\documentclass[a4paper,11pt]{article}
\pdfoutput=1 

\usepackage{jcappub} 
\usepackage[T1]{fontenc} 

\usepackage[colorlinks=true,linkcolor=blue]{hyperref}
\usepackage{color}
\usepackage{cancel}

\definecolor{mypink1}{rgb}{0.858, 0.188, 0.478}
\definecolor{mypink2}{RGB}{219, 48, 122}
\definecolor{mypink3}{cmyk}{0, 0.7808, 0.4429, 0.1412}
\definecolor{mygray}{gray}{0.6}
\definecolor{pptbg}{rgb}{0.961,0.945,0.863}

\newcommand{\be}[1]{\begin{equation} \label{#1}}
\newcommand{\ee}{\end{equation}}
\newcommand{\bea}{\begin{eqnarray}}
\newcommand{\eea}{\end{eqnarray}}
\newcommand{\ba}{\begin{array}}
\newcommand{\ea}{\end{array}}
\newcommand{\bel}{\begin{align}}
\newcommand{\eel}{\end{align}}
\newcommand{\nn}{\nonumber}

\newcommand{\tcr}{\textcolor{red}}

\newcommand{\oo}{$\mathcal{O}$ }

\newcommand{\PE}{$\mbox{P}_{\rm E}$}
\newcommand{\PT}{$\mbox{P}_{ b}$}
\newcommand{\PB}{$\mbox{P}_{\rm R}$}
\newcommand{\PM}{$\mbox{P}_{\rm M}$}

\title{\boldmath Spherically Symmetric Wormholes with anisotropic matter}

\author[a]{Hyeong-Chan Kim,}
\author[a,1]{Youngone Lee\note{Corresponding author.}}

\affiliation[a]{School of Liberal Arts and Sciences, Korea National University of Transportation, Chungju 380-702, Korea}

\emailAdd{hyeongchan@gmail.com}
\emailAdd{youngone@ut.ac.kr}

\abstract{
We study the geometry of a wormhole spacetime filled with anisotropic matter
in the context of general relativity.
In the course of the study, new static and spherically-symmetric solutions, analytic and numerical ones, are found.
We specify the existence condition for a wormhole throat.
We analyze properties of the solutions after categorizing them based on the spacetime regularity and the signature of energy density. 
The necessary conditions which allow a wormhole spacetime to be nonsingular are described.
}

\begin{document}
\maketitle
\flushbottom

\section{Introduction}\label{sec:intro}

Wormhole spacetimes are solutions of general relativity
~\cite{Morris:1988cz,Morris:1988tu,Kim:2001ri,Sushkov:2005kj,Lobo:2005us,Bronnikov:2017kvq}
which differ from other solutions such as massive stars and black holes.
The key difference lies in the topology. 
Two asymptotic regions are connected by a passage called a `throat'.
Since general relativity is a local theory,
it does not tell what the topology of whole spacetime should be
and does not exclude wormhole spacetimes.

Although wormholes are genuine solutions of general relativity,
the study of wormholes had not drawn much interest of researchers until recently.
The main reason is that
the existence of wormhole requires
violation of energy conditions, which are strong constraints to 
matter 
based on human intuitions on 
  ordinary classical matter.
To pass a throat of a wormhole, 
the geodesics of light bundle converge and then diverge
which requires `exotic' or `phantom' matter 
having negative pressure~\cite{Morris:1988tu,Raychaudhuri:1953yv}.
Another reason is that
this unique topology obscures 
many previously accepted physical concepts.
One may construct a time machine~\cite{Morris:1988tu,Frolov:1990si},
may have `charge without charge'~\cite{Misner:1957mt}
and 
the ADM mass measured in one side of the wormhole
may not be identical to that in the other side
~\cite{Frolov:1990si,Kim:2019elt}.

Despite of this pathology,
researchers in this field have studied wormholes as  physical objects
in the hope of future resolution of energy conditions.
Recently, an attempt to construct `phantom free models'
was suggested~\cite{Bronnikov:2018uje}
while others  have tried to get around the pathology
by constructing a wormhole using `ordinary' matter 
in modified theories of gravity
~\cite{Mehdizadeh:2015jra,Zangeneh:2015jda,Mazharimousavi:2016npo,Nandi:1997mx,Eiroa:2008hv}.
On the other hand, in quantum world, 
many examples were found which allow the violation of energy conditions.
Casimir effect~\cite{Casimir:1948dh}, 
negative energy region of squeezed light~\cite{Ford:2000ck}
 and the dark energy 
are the most prominent examples.
Therefore,
the violation of the energy condition is acceptable
for such special circumstances.

Recently, wormholes have  drawn interests
as a test-bed to understand quantum entanglement.
A non-traversable wormhole is regarded as a pair of entangled black holes
in `ER=EPR' conjecture
to understand the information paradox~\cite{Maldacena:2013xja}.
Traversable wormholes have also been suggested 
to have physically sensible interpretation of the conjecture~\cite{Maldacena:2018gjk}.
The study of wormholes as physical entities
in understanding the quantum-gravity
is getting more important than ever at this juncture.

Our purpose in this paper is, 
as a preliminary step, 
to categorize  wormhole solutions
and study the characteristic features of those solutions
to attain more clear grasp of the physical aspects of wormholes.
We focus on static, spherically symmetric traversable wormholes 
 which consist of anisotropic matter in the context of general relativity.
The anisotropic matter we are considering here
satisfies linear equation of state
throughout the entire space.
Hence isotropic perfect fluid
is a special case  of our study.
Although the anisotropic matter is distributed throughout the entire space,
one can obtain a spacetime of desirable asymptotic properties
or of minimal `exotic' matter
with appropriate junction conditions to match a truncated solution of us~\cite{Israel:1966rt}.

Traditionally  wormhole solutions have been obtained 
after assuming a well-behaving wormhole metric first
then calculating the Riemann tensor 
to get an appropriate stress-energy tensor
for matter fields afterward~\cite{Visser:1995cc}.
On the other hand, in this work,
we begin with the specification of matter fields
satisfying a given equation of state 
and then solve Einstein's equation.
After that, we find out the condition for  
a wormhole solution to be nonsingular over the whole spacetime.
We also analyze  properties
satisfied by the general solutions
and categorize them into six types 
based on the spacetime regularity and the signature of energy density.

The existence condition for a wormhole throat and
 properties of Einstein equation are discussed in Sec.~\ref{sec:condition}.
Analysis of general solutions is in Sec.~\ref{sec:general}
where behavior of solutions around special points such as
points at asymptotes, the throat and the repelling point
are analyzed.
Numerical solutions are also presented in Sec.~\ref{sec:numeric}
and a few exact solutions are presented in Sec.~\ref{sec:exact}.
Properties of a few  limiting solutions
are presented in Sec.~\ref{sec:slimit}.
Finally, 
we summarize the results 
and discuss the properties of the solutions
and suggest future works to be done
in Sec.~\ref{sec:summary}.

\section{General Properties and Existence Conditions}\label{sec:condition}
For simplicity, we consider  static and spherically symmetric configurations.
The stress-energy tensor for an anisotropic fluid compatible with the spherical symmetry is given by
\be{st}
T^{\mu\nu} = \rho u^\mu u^\nu + p_1 r^\mu r^\nu 
+ p_ 2 (\theta^\mu \theta^\nu+\phi^\mu \phi^\nu),
\ee
where $\rho$ is the energy density measured by a comoving observer with the fluid.
The vectors $u^\mu,r^\mu,\theta^\mu$ and $\phi^\mu$ are mutually orthogonal
and denote four-velocity, radial and  the unit angular vectors, respectively.
As we mentioned above, 
the radial and the angular pressures are assumed to be proportional to the density:
\be{eos}
p_1 = w_1 \rho, \qquad p_2 =	 w_2 \rho,
\ee
with constant $w_1$ and $w_2$.
We focus on a static  and spherically symmetric configuration
of which the line element\footnote{Because we deal with spherically symmetric situation, 
many parts of the calculation in this section overlap with those in Ref.~\cite{Kim:2017hem}.
Therefore, we leave only the main results.}
 is given by

\be{metric}
ds^2 = - f(r) dt^2 + \frac{1}{1-2m(r)/r} dr^2 + r^2 d\theta^2 + r^2 \sin^2\theta d\phi^2.
\ee

\subsection{Tolmann-Oppenheimer-Volkhoff equation}
The $G_{tt}$ part of the Einstein equation defines the mass function $m(r)$ as,   
\be{gr}
 m(r) = 4\pi \int^r r'^2 \rho(r') dr',
\ee
where  an  integration constant is absorbed into the definition of $m(r)$.
The continuity equation $\nabla^\mu T_{\mu\nu} =0$ yields
 the  Tolmann-Oppenheimer-Volkhoff (TOV) equation for an anisotropic matter,  
\be{TOV}
p_1' = -(\rho + p_1) \frac{m+ 4\pi r^3 p_1}{r(r-2m)} + \frac{2(p_2 - p_1)}{r}.
\ee
Here the prime means to take the derivative with respect to $r$.

Then $g_{tt}$ part of the metric, or $f(r)$, 
can be obtained in two different ways. 
Combining the relation $G_{tt}=8\pi T_{tt}$  and $G_{rr}=8\pi T_{rr}$ 
we have
\bea
\label{gtt}
\frac{f'}{f} = \frac{2(m + 4\pi r^3 p_1)}{r(r-2m)} .
\eea
The equation can be directly integrated to give
\be{fr:g}
 f(r) =  \tilde f_0 \frac{ (r-2m)^{-w_1}}{r }  \exp \left[(1+w_1) \int_{r_0}^r \frac{1}{r- 2m(r)} dr\right],
\ee
where $\tilde{f}_0$ is an integration constant. 
The $w_2$-dependence
will be restored  in \eqref{fr:g} when $m(r)$ is expressed explicitly.
The sign of $\tilde f_0$ will be chosen
so that the signature of the metric be Lorentzian
because the Einstein equation does not determine the sign of $f(r)$.
On the other hand,
the anisotropic TOV equation \eqref{TOV}  with
 the equation of state~\eqref{eos}
reads
$$
\frac{\rho'}{\rho} = -\frac{1+w_1}{2w_1} \frac{f'}{f} 
	+ \frac{2(w_2-w_1)}{w_1 }\frac{1}{r}.
$$
After integrating one gets
\be{f:r1}
f(r) =f_b \Big(\frac{r}{b}\Big)^{\frac{4(w_2-w_1)}{1+w_1}} 
	\Big(\frac{\rho}{\rho_b}\Big)^{-\frac{2w_1}{1+w_1}} ,
\ee
where $b$ is the value of the radial coordinate at the throat.
Here, $\rho_b$ and $f_b$ are the values of energy density
and the function $f$ at the throat.
We have two forms of $f(r)$, \eqref{fr:g} and \eqref{f:r1}, and
both are useful for later considerations.

The remaining task is to find the explicit form of $m(r)$
by solving the TOV equation\footnote{When $w_1=-1$ and $w_1=-1/3=w_2$, 
the TOV equation allows exact solutions 
as in Ref.~\cite{Cho:2017nhx,Cho:2016kpf}. }.
Using Eqs.~\eqref{eos} and~\eqref{gr}, the TOV equation becomes
\be{TOV2}
\frac{m''}{m'} = -\frac{1+w_1}{2w_1} \frac{ 1 
	+2w_1 m'}{r-2m} +\frac{1+w_1+4w_2}{2w_1 r},
\ee
where we use $\rho' /\rho = m''/m' - 2/r$.
Equation~\eqref{TOV2} does not allow an analytic solution in general.
However, the above relation can be cast into
 a first order autonomous equation in a two dimensional $(u,v)$ plane: 
\be{de2}
\frac{du}{dv} 
 = \frac{-1}{1+w_1} \frac{(1-u)(2v-u)}{v\left[ v- v_b
 	+ s (1-u)\right]},
\ee
where $u$ and $v$ are scale-invariant variables defined by
\be{uv}
u \equiv \frac{2m(r)}{r}, \qquad v \equiv \frac{dm(r)}{dr} =4\pi r^2 \rho .
\ee
The constants $v_b$ and $s$ denote
 the value of $v$ at the wormhole throat 
and the {\it slope} of the line $v-v_b + s(1-u) =0$, respectively,
\be{slope}
v_b\equiv -\frac1{2w_1}, \qquad  
s \equiv -\frac{1+w_1+ 4w_2}{2w_1 (1+ w_1)}.
\ee
Given $u$ and $v$, 
the metric $g_{tt}$ can be reconstructed from $v$ via Eq.~\eqref{f:r1} 
and $g_{rr}$ from $u$ respectively.

\subsection{The existence condition for a wormhole throat}
\label{thcond}

The geometry of  each side of the throat
 is described by the metric~\eqref{metric} with an appropriate mass function.
For the metric~\eqref{metric} to have a throat at $r=b$, 
its spatial part should have a coordinate singularity of the form 
$g_{rr} \approx g (1-b/r)^{-1}$ with $g>0$ where $r$ 
is an areal radial coordinate\footnote{
Consider $g_{rr} \approx g(1-b/r)^{-\beta}$.
When $\beta <1$ and $\beta\neq 0$, 
the geometry must be singular~\cite{Kim:2017hem}.  
When $\beta \geq 2$, the geometry at $r=b$ can be regular 
but the surface $r=b$ is located at infinity implying that the throat is non-traversable. 
Therefore, to be traversable, the value should satisfy $1\leq \beta < 2$.
However, when $\beta$ is non-integer, higher curvatures must have a singularity at $r=b$. 
In this sense, we restrict our interest to $\beta =1$ case to find a nonsingular traversable wormhole. 
}.
We then, from Eq.~\eqref{fr:g}, find
$$
f(r) \propto \frac{g^{w_1} }{r} (r-b)^{(1+w_1)g -w_1}  .
$$
At the wormhole throat, $f(b)$ should take a finite value,
hence the value of $g$ becomes
$$
g \equiv \lim_{r\to b} \left(1-\frac{b}{r}\right) g_{rr} = \frac{w_1}{1+w_1} > 0.
$$
Therefore, a wormhole solution exists 
only when $w_1< -1$ or $w_1>0$.\footnote{When $w_1 = 0$ and $g=0$, 
the spatial geometry does not form a throat at $b$. 
The solution for $w_1=-1$ was analyzed in Ref.~\cite{Cho:2017nhx} thoroughly
in which no wormhole-like solutions were found.} 
Around the throat, the mass function $m(r)$ behaves as,
from $g_{rr}=(1-2m(r)/r)^{-1}$,
\be{vT}
m(r) \simeq \frac{b}{2w_1}\left(1+w_1- \frac{r}{b}\right) \quad
\Rightarrow \quad  m'(b)= -\frac{1}{2w_1}.
\ee
From this, one can notice that
 the value of $v\equiv m'(r)$ at the wormhole throat is solely determined by $w_1$\footnote{
One can confirm that all three solutions in Ref.~\cite{Morris:1988cz} satisfy this relation.
For the case that $w_1$ depends on $r$, $w_1=w_1(r)$ as in Ref.~\cite{Halder:2019urh},
one can show that $m'(b)=-1/2w_1(b)$.
}.
Previous definition of $v_b(=-1/2w_1)$ comes from this observation.

Based on the observation above, let us divide the wormhole `throats' into two types:
\begin{enumerate}
\item When $w_1>0~ (v_b<0)$: the mass function decreases  from $b/2$ with $r$,
meaning that the mass in the region $(b, R> b)$ is $\Delta m=m(R)-m(b)<0$.
The energy density at the throat takes a negative value $\rho(b)= v_b/4\pi b^2< 0$.

\item When $w_1< -1~(0<v_b < 1/2)$:
the mass function increases with $r$ and the density takes a positive value.  
Therefore, if we use matter with positive energy density to weave a wormhole throat, 
we cannot avoid phantom-like matter satisfying $w_1< -1$.

\end{enumerate}

As we  have commented previously, 
we consider nonsingular solutions 
in which all quantities are continuous
throughout the entire spacetime.
If necessary,
solutions with a surface layer or
discontinuous energy of a hypersurface~\cite{Israel:1966rt}
can be obtained by cut and paste  our solutions
with appropriate junction conditions.
To express both sides of the throat continuously,
 one may introduce a new radial coordinate $x$ around the throat,
\be{x:r}
\sqrt{g_{rr}(r)} dr = \pm b dx \quad \Rightarrow \quad 
	\pm x \simeq 2\sqrt{g} \sqrt{\frac{r}{b} -1} ,
\ee
where we choose $x=0$ at the throat and $\pm$ for $x \gtrless 0$, respectively.
This gives 
$$
\frac{d m(x)}{dx} = \frac{dr(x)}{dx} m'(r) =  \frac{-(1+w_1)}{4w_1^2} x .
$$
Therefore, there is no singularity or discontinuity of density at the wormhole throat. 
Note that the existence condition for a wormhole throat is independent of the angular pressure. 

It is to be noted that 
the radial coordinate does not have to be areal
to show the inevitable use of `exotic' matter.
As in Ref.~\cite{Bronnikov:2018vbs}, 
the violation of energy conditions can also be derived
by requiring  $r(x)$ to be a minimum at the throat 
with respect to the regular $x$-coordinate
combining with the Einstein equations.
The use of areal radial coordinate here
enables us to show 
the independence of the angular pressure 
at the throat threaded by anisotropic matter.
In other words,
the angular pressure can have any value
that makes the solution be nonsingular.

\section{Analysis for general solutions}\label{sec:general}

The solution to the autonomous equation~\eqref{de2} 
can be represented
by an integral curve $C$ on the two dimensional plane $(u,v)$.
Hence, to sketch general properties of solutions 
we need to closely look at the autonomous equation.

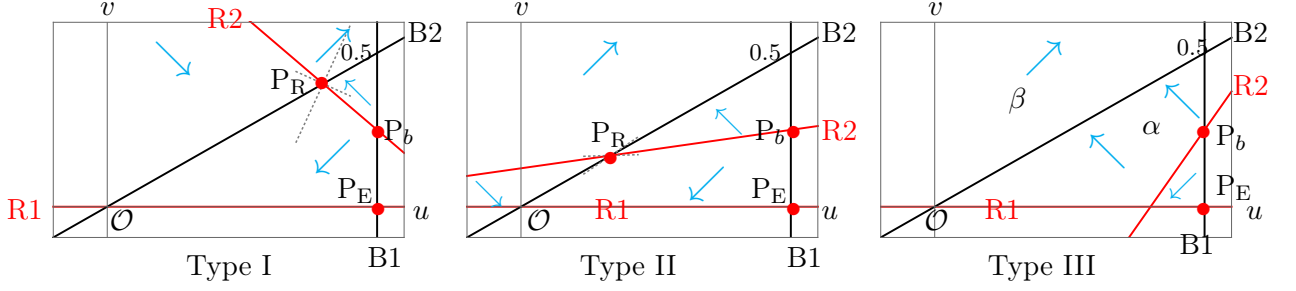
\begin{figure}[htb]
\begin{center}
\begin{tabular}{ccc}
\includegraphics[width=.3\linewidth,origin=tl]{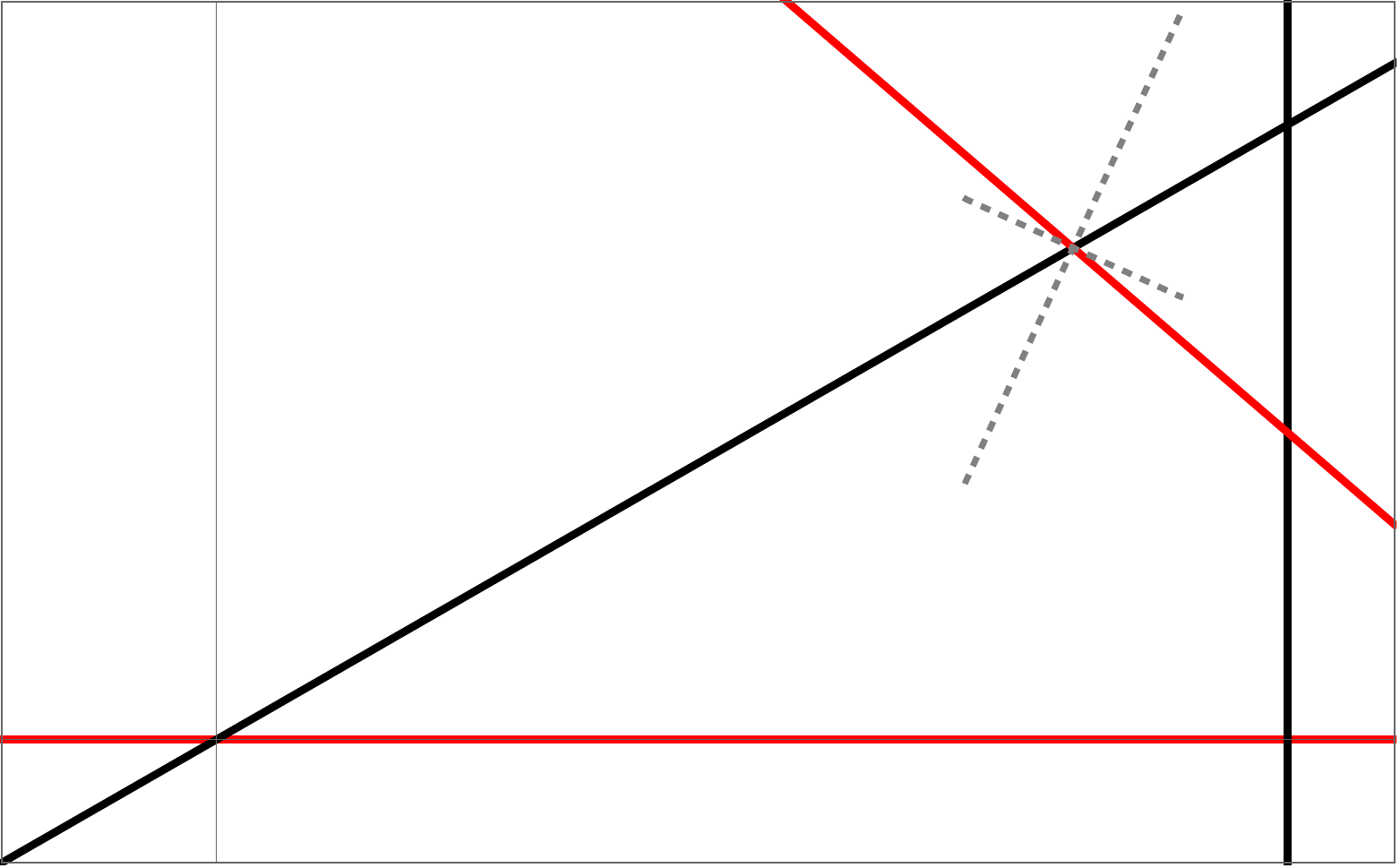}
&\quad
\includegraphics[width=.3\linewidth,origin=tl]{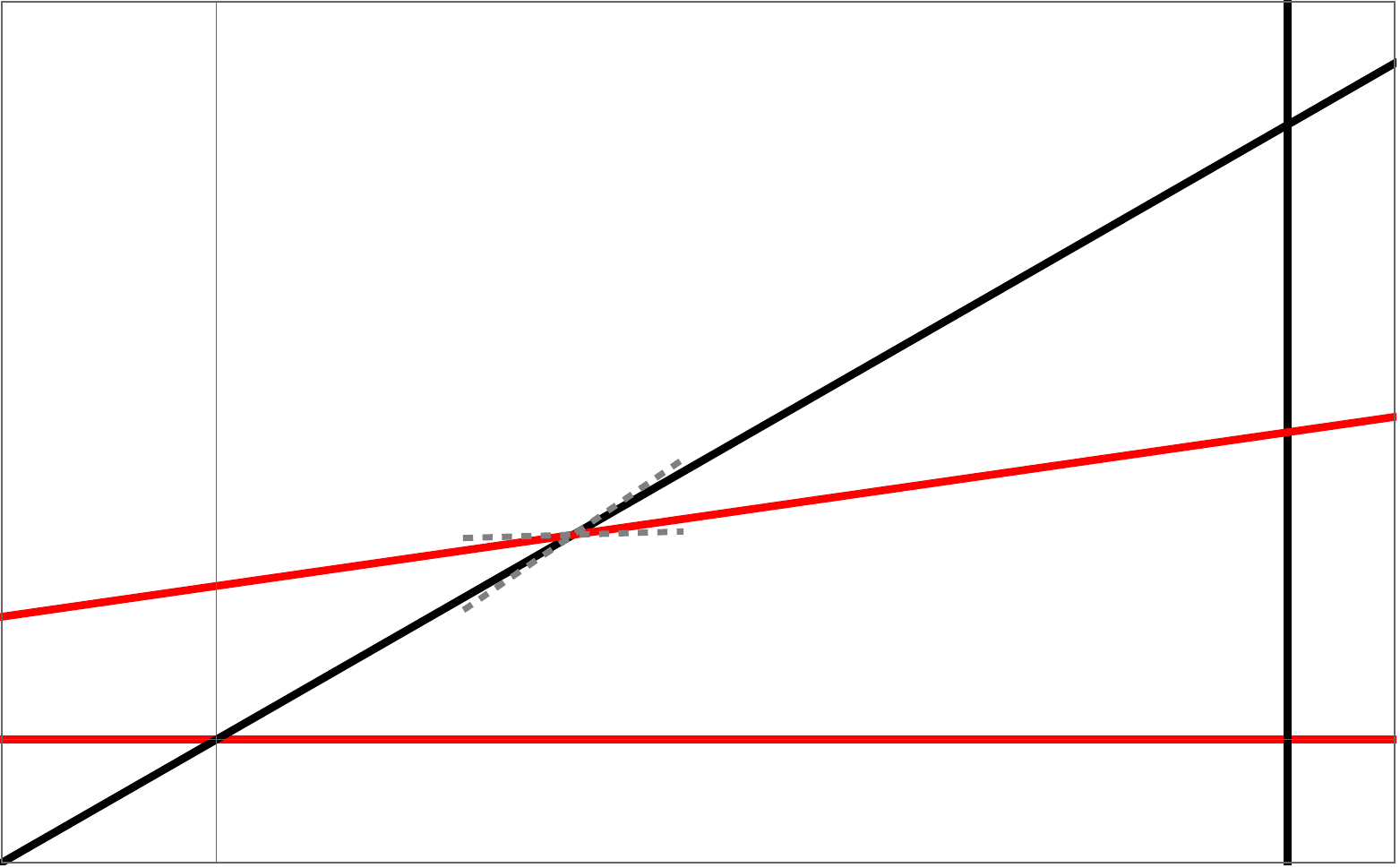} 
&\quad
\includegraphics[width=.3\linewidth,origin=tl]{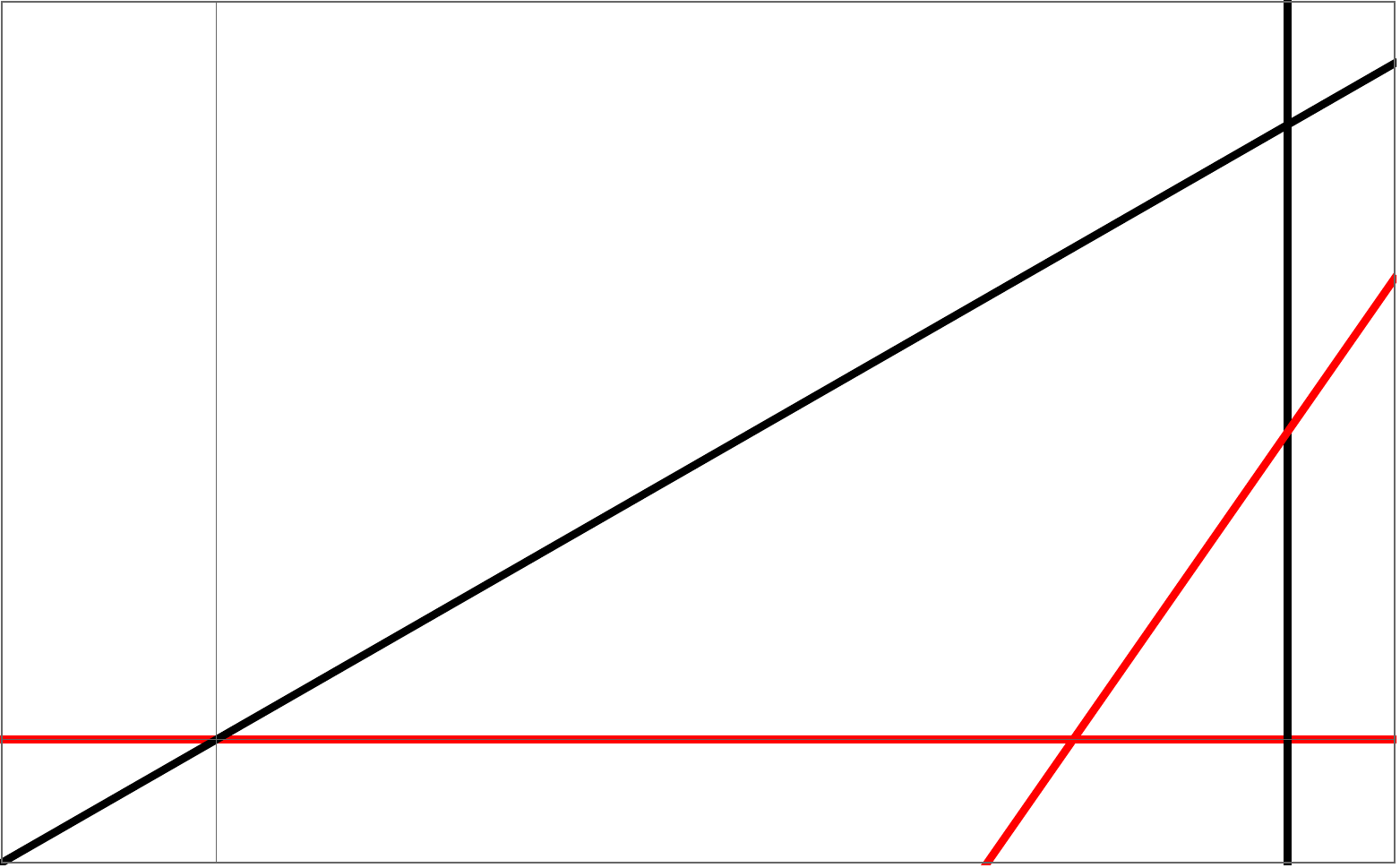}
\\
Type I & Type II  &  Type III
\end{tabular}
\put (-313,-22) {$u$  }
\put (-430,55) {$v$  }
\put (-430,-26) { $\mathcal{O}$  }
\put (-315, 45) {B2}
\tcr{\put (-465, -22) {R1}
\put (-390, 50) {R2}}
\put (-330, -40) {B1}
\put (-324,8) {\PT} \put (-329,8) {\tcr{\large $\bullet$}}
\put (-345,-14) { \PE} \put (-329,-21) {\tcr{\large $\bullet$}}
\put (-370,27) { \PB} \put (-350,26.5) {\tcr{\large $\bullet$}}
\put (-340,38) {\footnotesize{0.5} }
\put (-410, 35) {\textcolor{cyan}{ \Large $\searrow $  }}
\put (-339, 23) {\textcolor{cyan}{  $\nwarrow $  }}
\put (-350, -2) {\textcolor{cyan}{ \Large $\swarrow $  }}
\put (-350, 40) {\textcolor{cyan}{ \Large $\nearrow $  }}
\put (-160,-22) {$u$  }
\put (-275,55) {$v$  }
\put (-274,-26) { $\mathcal{O}$  }
\put (-160, 45) {B2}
\tcr{\put (-245, -22) {R1}
\put (-160, 8) {R2}}
\put (-173, -40) {B1}
\put (-185,8) {\PT} \put (-173.5,8) {\tcr{\large $\bullet$}}
\put (-190,-14) { \PE} \put (-173.5,-21) {\tcr{\large $\bullet$}}
\put (-250,7) { \PB} \put (-242,-2) {\tcr{\large $\bullet$}}
\put (-190,38) {\footnotesize{ 0.5} }
\put (-290, -15) {\textcolor{cyan}{  $\searrow $  }}
\put (-250, 35) {\textcolor{cyan}{ \Large $\nearrow $  }}
\put (-210, -12) {\textcolor{cyan}{ \Large $\swarrow $  }}
\put (-200, 12) {\textcolor{cyan}{ $\nwarrow $  }}
\put (-1,-22) {$u$  }
\put (-120,55) {$v$  }
\put (-125,-26) { $\mathcal{O}$  }
\put (-6, 45) {B2}
\put (-90, 20) {$\beta$}
\put (-40, 10) {$\alpha$}
\tcr{\put (-99, -22) {R1}
\put (-6, 25) {R2}}
\put (-26, -35) {B1}
\put (-16,6) { \PT} \put (-20,8) {\tcr{\large $\bullet$}}
\put (-16,-13) { \PE} \put (-20,-21) {\tcr{\large $\bullet$}}
\put (-30,40) {\footnotesize{ 0.5} }
\put (-84, 35) {\textcolor{cyan}{ \Large $\nearrow $  }}
\put (-60, 0) {\textcolor{cyan}{ \Large $\nwarrow $  }}
\put (-32, 19) {\textcolor{cyan}{ \Large $\nwarrow $  }}
\put (-30, -12) {\textcolor{cyan}{ $\swarrow $  }}
\end{center}
\caption{
Classification of the autonomous equation for $w_1< -1$. 
The cyan arrows represent the increasing direction of radial coordinate $r$.  
The direction changes based on the lines B2 and R2.
The red line R2 changes depending on the values of $w_i$.
}
\label{fig:cases}
\end{figure}

There are four interesting lines on which 
the denominator or the numerator is equal to zero
or equivalently, $dv/du=0$ or $du/dv$=0.
These four lines 
are plotted in Figs.~\ref{fig:cases} and \ref{fig:cases2}.
We call the thick-red lines R1 and R2
on which only $u$ changes its value ($dv/du=0$) and 
$C$ crosses the lines horizontally:
\be{Rline}
\mbox{R1:} ~v=0, \qquad \mbox{R2:} ~ v -v_b =s(u-1).
\ee
Energy density $\rho\propto v$  vanishes on the line R1.
Because R1 is horizontal, solution curves
are allowed to touch R1 only at a special points $\mathcal{O}(0,0)$ and \PE$(1,0)$
because $dv/du=0$ on R1.

We call the thick-black lines B1 and B2 
on which only $v$ changes its value ($du/dv=0$) and 
$C$ crosses the lines vertically:
\be{Bline}
\mbox{B1:}~ u=1, \qquad \mbox{B2:} ~u=2v .
\ee 
The line B1 represents a static boundary where $g_{rr}$ changes its signature.
Because the line B1 is parallel to the $v$ axis, 
$C$ is allowed to cross the line only through \PE, 
where the subscript $E$ denotes event horizon.
Since we focus on  static traversable wormhole solutions in this work, 
we restrict our interests to the region with $u \leq 1$. 
As will be shown later in this work,
a solution having a traversable wormhole throat does not have an event horizon,
 i.e., its solution curve never crosses B1.\footnote{
If the equation of state is nonlinear,
or $w_1,w_2$ are functions of $r$ 
as in Ref.~\cite{Bronnikov:2017kvq},
a wormhole spacetime 
having  a cosmological horizon could exist.
At the present work,
any solution curve  having a wormhole throat,
does not go into the time dependent region.
In other words,
the sign of temporal component of the metric, $g_{tt}$,
does not change here(For details see footnote~\ref{foot7}).
To be specific,
we consider the theory of general relativity
without cosmological constant
and matter with constant equation of states $w_1$ and $w_2$ satisfying
 $w_1<-1$ or $w_1>0$.
}
Note that solution curves never become vertical or horizontal at the points 
which are not on the lines R1, R2, B1, and B2.

The crossing points between the black lines and the red lines are 
 \be{crossing}
\mathcal{O}(0,0), \quad \mbox{\PE}(1,0), \quad 
\mbox{\PT}\big(1, v_b \big), \quad
\mbox{\PB} (2v_R, v_R),
\ee
where $ v_R\equiv 2w_2/[(1+w_1)^2+ 4w_2].$
The point \oo plays the role of an asymptotic geometry of $r\to \infty$ for the wormhole spacetime.
The point \PT ~is a wormhole throat.  
The point \PB~is a point where B2 meets R2
and acts like a `repeller' as described in Sec.~\ref{sec:PB}.
Because solutions tend to avoid this point, 
we call \PB~`Repelling point'.
The radial dependence of a solution curve $C$ on the $(u,v)$ plane 
 can be obtained from 
\be{dr:u}
\frac{dr}{r} = \frac{du}{2v-u} = \frac{-1}{1+w_1} \frac{1-u}{v[v-v_b + s(1-u)]} dv,
\ee
where the second equality comes from Eq.~\eqref{de2}.
The relation can be interpreted as:
\begin{itemize}
\item The radius increases/decreases with $u$ 
when the curve is above/below the line B2, respectively.
\item The radius increases or decreases with $v$ 
depending on many factors.
They are the signature of $1+w_1$ and
whether the curve is above or below the lines R1 and R2.
\end{itemize}
Hence, the region $(u<1,v)$ is 
 divided into 7 regions maximally 
by the 3 lines, R1, R2 and B2.
The increasing direction of $r$ varies
region by region.
The `cyan' arrows in the figures in this work
 represent the increasing direction of $r$.
One can intuitively notice
how the solution (mass and density) develops 
as one moves from the throat to the outside regions.

As an example,
let us apply the above properties 
to the Type I and II of the Fig.~\ref{fig:cases}.
Consider a solution curve which passes \PT~vertically.
If it goes vertically upwards from \PT,
the solution curve $C$ bends to the left 
following the cyan arrow 
until the curve meets the line B2 or R2.
If it touches the line B2,
the curve bends upwards and approaches the line B1 indefinitely.
This asymptotic behavior with
$v=m'(r)\to\infty$ can be excluded 
as a physically viable regular solution.
By applying the same consideration,
one can figure out genuine regular solutions.

\subsection{Analysis for $w_1< -1$ and $\rho>0$}
Let us first consider the case with $w_1< -1$. 
In this case, $v_b>0$. 
In terms of the slope $s$ of the line R2, 
we divide the deployment of the points and the lines into three different types as in Fig.~\ref{fig:cases},
\be{cases}
\mbox{I:} ~ s < 0 ~~ \left(w_2 > -\frac{1+w_1}{4} \right) , \quad 
\mbox{II:} ~ 0 \leq s  \leq v_b~~ \left(0 \leq w_2  \leq -\frac{1+w_1}{4} \right) , \quad
\mbox{III:}~ s >  v_b  ~~ (w_2 < 0).\nn
\ee
For the Types I and II the point \PB~is 
in the region we are interested in($0\leq u< 1$). 
For the Type III, \PB~is located outside of the  region.
In Sec.~\ref{sec:exact}, we present and display exact solutions which belong to the types I and  II.

Even though there is a wormhole throat,
the solutions in the Type III does not  have a regular asymptotic region:
Most of the solution curves allowing 
a wormhole throat pass the point \PT~with $u\leq 1$
and tangent to the line B1 at the throat except a symmetric solution described in Sec.~\ref{sec:332}.
Let us consider one of the curves.
A solution curve may depart from \PT~vertically upwards (increasing $v$, decreasing $u$)
and downwards (decreasing $u$ and $v$).
As one can see clearly from the arrows,
the first one that starts from the region $\alpha$
 continues until it meets the line B2
and becomes vertical
followed by  entering the region $\beta$.
Once a solution curve enters the region $\beta$
it goes upwards 
to a singular region $v\to\infty$
until the curve becomes parallel to the line B1 ($u=1$).
The second one touches R2 at some point\footnote{
\label{foot7}
If the curve $C$ touches R1, it is possible only at \PE~in principle.
This is not possible because the autonomous equation implies 
that the solution is analytic.
Putting the trial function $u=1- \kappa |v|^\beta $ with $\kappa |v|^\beta \ll 1$ to Eq.~\eqref{de2} we get $
 \beta = (-2w_1)/(1+w_1) < 0.$
Therefore, the solution curve $C$ never meet the point \PE~but bounces for any value of $\kappa \neq 0$. 
This implies that event horizons will never be formed. 
},
 and becomes horizontal there
entering to the region $\alpha$.
In this region, $v$ cannot decrease but only increases
until $C$  meets the line B2.
There, $C$ becomes vertical again 
and enters the region $\beta$
and repeats the same behavior as above.
In summary, both ends of $C$ give infinity value of $v$, implying singularities.
Therefore, later in this work, we consider only the type I and II with $w_2 \geq 0$.

\subsection{Analysis for $w_1 > 0$ and $\rho<0$}

\begin{figure}[htb]
\begin{center}
\begin{tabular}{ccc}
\includegraphics[width=.3\linewidth,origin=tl]{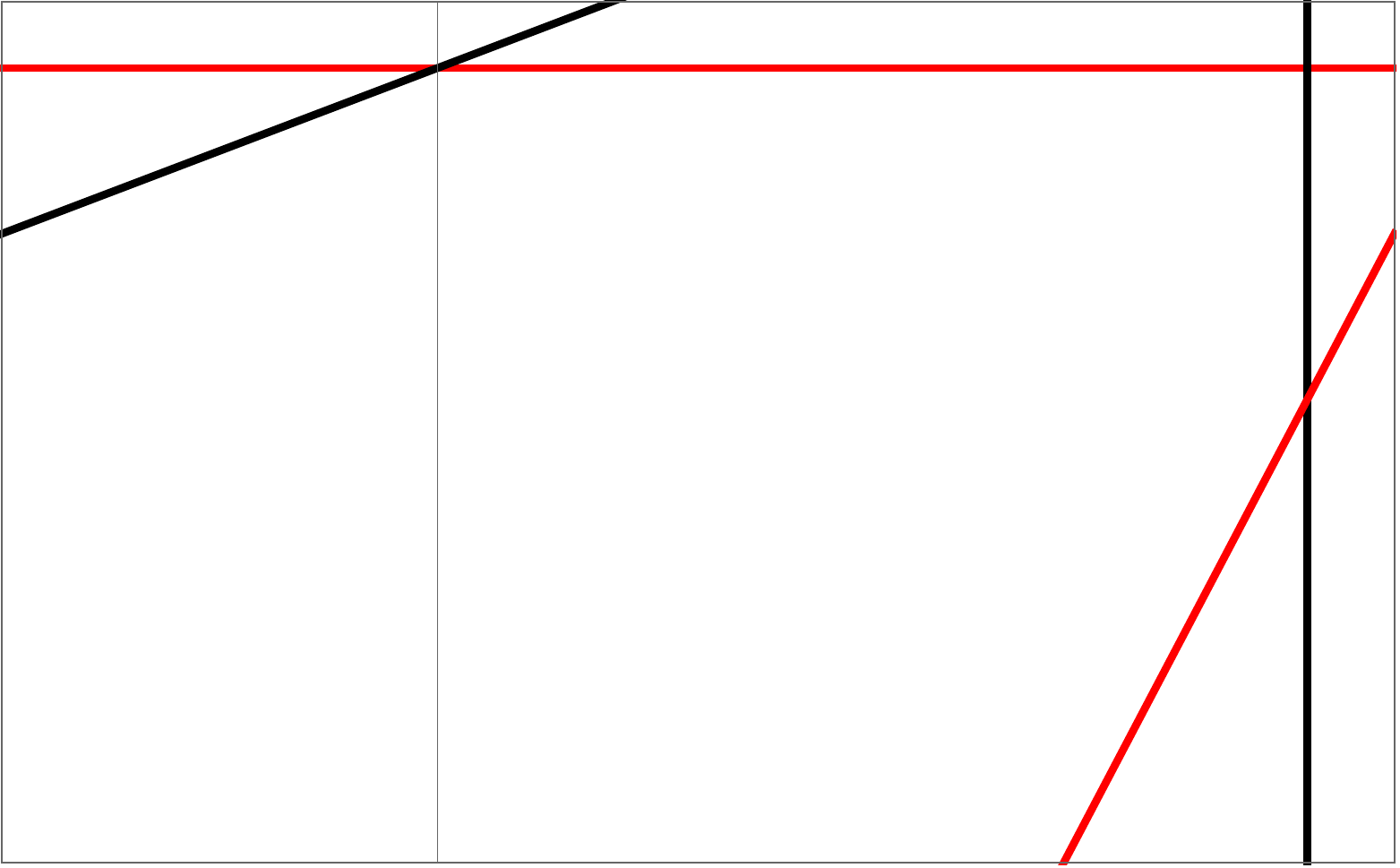}
&
\quad
\includegraphics[width=.3\linewidth,origin=tl]{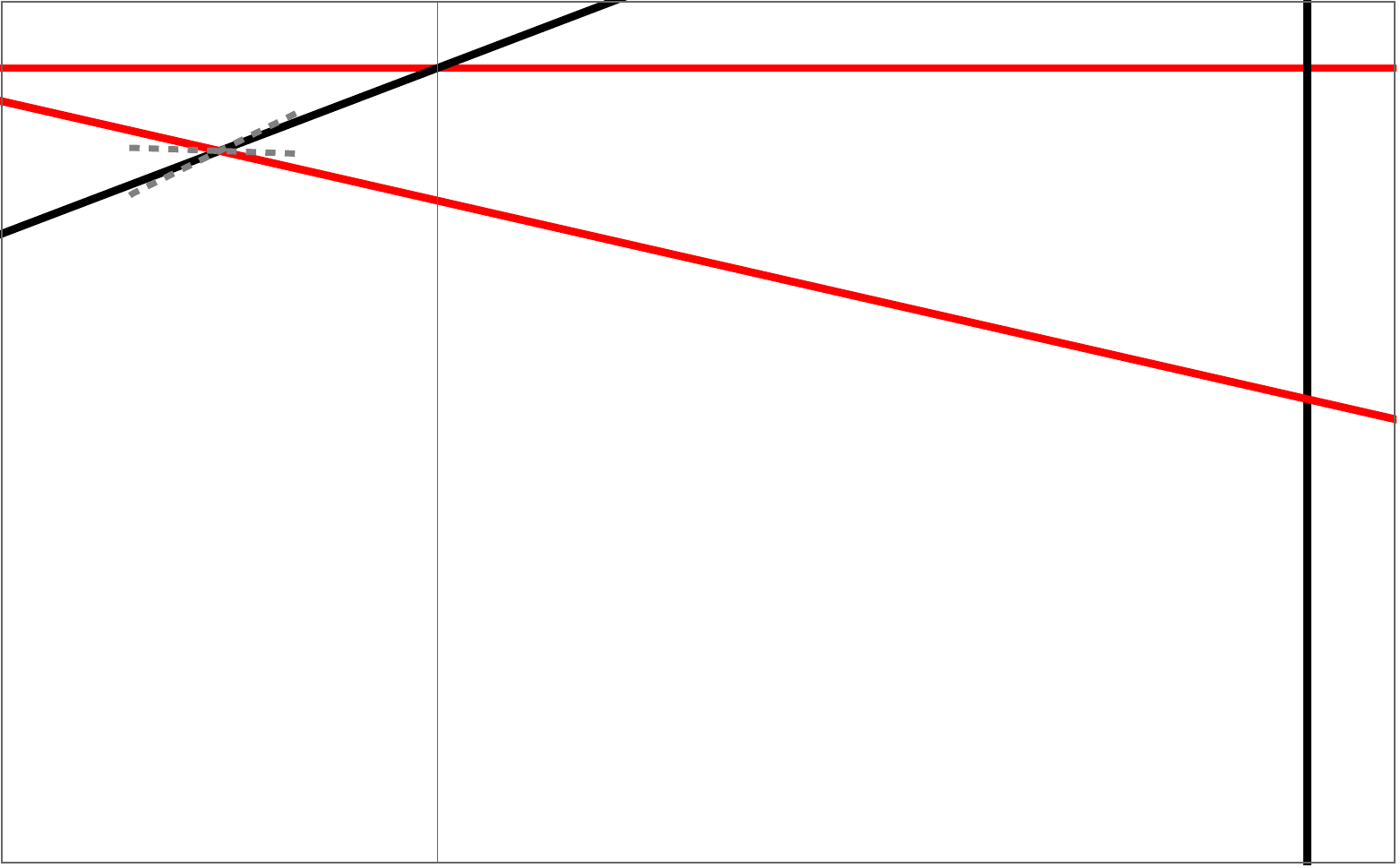}
&
\quad
\includegraphics[width=.3\linewidth,origin=tl]{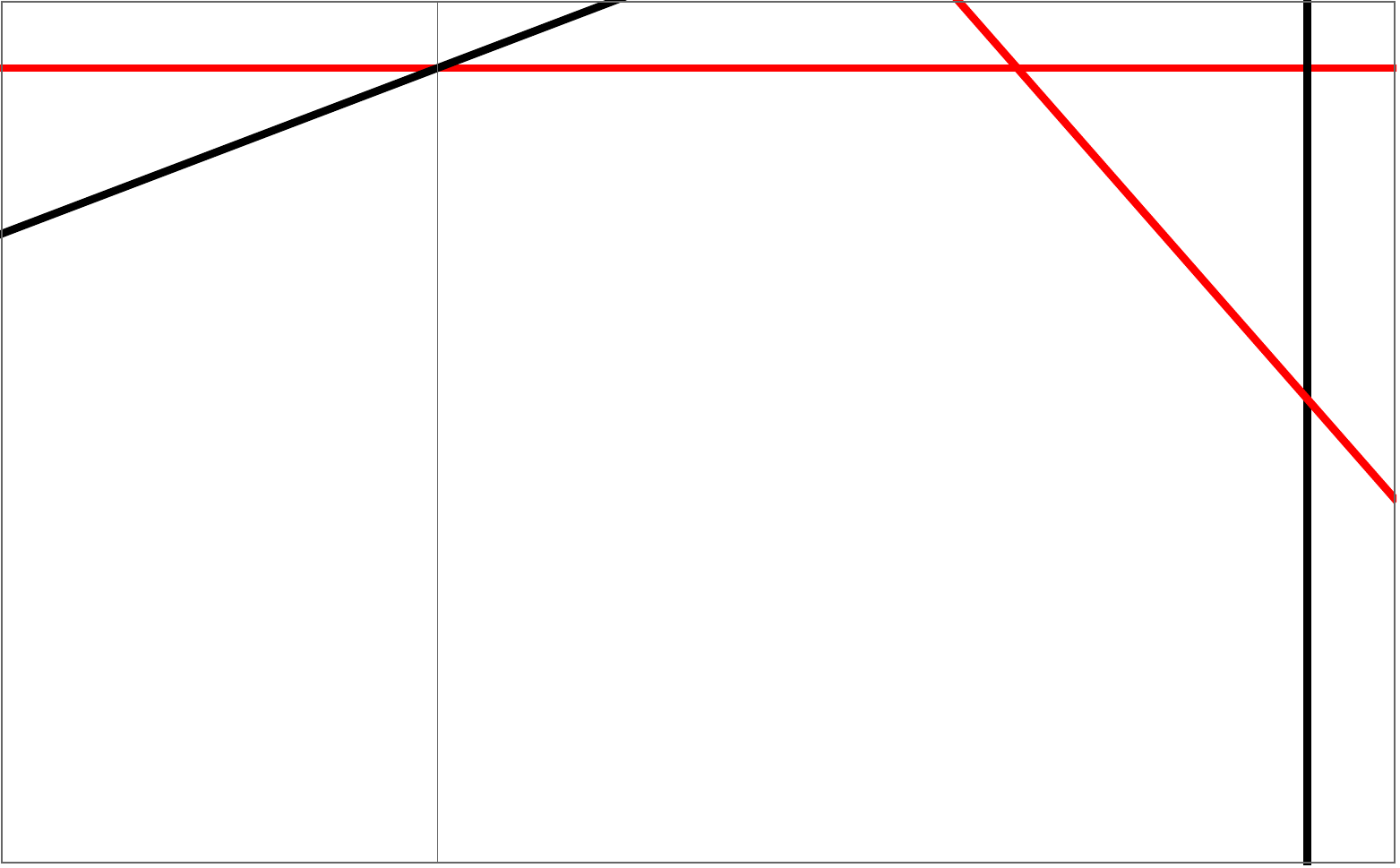}
\\
Type IV & Type V &  Type VI
\end{tabular}
\put (-333,47) {$u$  }
\put (-410,55) {$v$  }
\put (-410,35) { $\mathcal{O}$  }
\put (-465, 30) {B2}
\tcr{\put (-360, -22) {R2}
\put (-360, 50) {R1}}
\put (-330, -40) {B1}
\put (-343,10) {\PT} \put (-327,12) {\tcr{\large $\bullet$}}
\put (-343, 37) { \PE} \put (-327,42) {\tcr{\large $\bullet$}}
\put (-440, 52) {\textcolor{cyan}{  $\searrow $  }}
\put (-440, 38) {\textcolor{cyan}{ $\nearrow $  }}
\put (-370, 28) {\textcolor{cyan}{ \Large $\nwarrow $  }}
\put (-405, -12) {\textcolor{cyan}{ \Huge $\nwarrow $  }}
\put (-340, -20) {\textcolor{cyan}{ \Large $\swarrow $  }}
\put (-168,47) {$u$  }
\put (-255,55) {$v$  }
\put (-250,35	) { $\mathcal{O}$  }
\put (-305, 28) {B2}
\put (-277, 25) {\PB}
\tcr{\put (-209, 24) {R2}
\put (-220, 50) {R1}}
\put (-178, -40) {B1}
\put (-186,10) {\PT} \put (-173,12) {\tcr{\large $\bullet$}}
\put (-191, 40) { \PE} \put (-173,42) {\tcr{\large $\bullet$}}
\put (-270, 50) {\textcolor{cyan}{ $\searrow $  }}
\put (-275, 41) {\textcolor{cyan}{\tiny $\nearrow $  }}
\put (-300, 36) {\textcolor{cyan}{ $\searrow $  }}
\put (-200, 30) {\textcolor{cyan}{ \Large $\nwarrow $  }}
\put (-240, -12) {\textcolor{cyan}{ \Huge $\swarrow $  }}
\put (-1,47) {$u$  }
\put (-100,55) {$v$  }
\put (-100,35) { $\mathcal{O}$  }
\put (-153, 28) {B2}
\tcr{\put (-70, 50) {R1}
\put (-44, 25) {R2}}
\put (-20, -40) {B1}
\put (-35,15) { \PT} \put (-17,12) {\tcr{\large $\bullet$}}
\put (-35,42) { \PE} \put (-17,42) {\tcr{\large $\bullet$}}
\put (-130, 50) {\textcolor{cyan}{ \ $\searrow $  }}
\put (-30, 30) {\textcolor{cyan}{ $\nwarrow $  }}
\put (-110, -5) {\textcolor{cyan}{ \Huge $\swarrow $  }}
\put (-40, -5) {\textcolor{cyan}{ \Huge $\swarrow $  }}
\end{center}
\caption{
Classification of the autonomous equation for $w_1> 0$. 
The direction changes based on the line B2, R1 and R2.
}
\label{fig:cases2}
\end{figure}
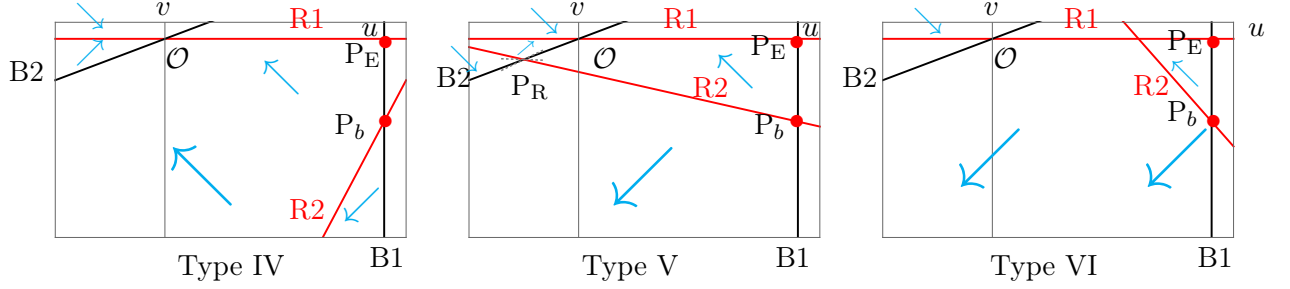
In this case, since $v_b < 0$,
the energy density near the wormhole throat   is negative.
There are three types:
\be{cases}
\mbox{IV:} ~s  > 0~~  \left(w_2  < -\frac{1+w_1}{4} \right) , \quad
\mbox{V:}~ v_b \leq s  \leq 0  ~~ \left(-\frac{1+w_1}{4} \le w_2 \le 0 \right) , \quad
\mbox{VI:} ~ s < v_b ~~ ( w_2 \geq 0).\nn
\ee

We first show that 
the Type VI does not allow a regular wormhole solution 
as in the Type III for $w_1<-1$.
Let us consider a solution curve $C$ passing \PT~vertically. 
Vertically upward part of the curve becomes horizontal when it meets the red lines 
R1 or R2.  
As can be understood in the figure, the curve will cross R2
before it meets $\mathcal{O}$ and becomes parallel to the $u$ axis.
Then, by similar consideration above (when $w_1<-1$),
the solution curve after crossing the line R2 
goes to the region $(u,v) \to (-\infty,-\infty)$. 
Vertically downward part of the curve at \PT~cannot be a parallel to the $v=0$ line 
because it never meet a red line. 
Again, the curve  goes to the region  $(u,v) \to (-\infty,-\infty)$.
In this case, a space-time singularity will form there. 

On the other hand, for the type IV, 
the vertically upward part of $C$ 
can approach $\mathcal{O}$,
hence some solutions follow the path.
The vertically downward part of $C$ can meet R2 
and becomes parallel and the value $v$ starts to increase. 
In that case, some curve can also approach the point $\mathcal{O}$.  
Therefore, there are possibilities
that both ends of a solution curve finish at $\mathcal{O}$, 
forming  regular asymptotic regions. 
Similar analysis for the type V leads that 
only upward curves have a possibility to approach $\mathcal{O}$,
to form a regular solution.

To summarize, we are mainly interested in the cases with Type I, II, IV, and V. 
For all the cases, $w_1 w_2 < 0$ is satisfied. 
Therefore, isotropic fluid fails to form a regular wormhole spacetime.

\subsection{Behaviors around important points}
To have clear pictures,
we study properties of solutions 
around a few important points.
One of them is \oo
which represents asymptotic limit ($r\to\infty$)
for the types I, II, IV and V.
Other important points 
are \PT~and \PB, 
representing a wormhole throat and the repelling point, respectively.

\subsubsection{The asymptotic behavior around \oo } \label{sec:O}
Let us first search for the behavior of $C$ around the point $\mathcal{O}(0,0)$.
Because $|u|, |v| \ll 1$ around $\mathcal{O}$, Eq.~\eqref{de2} can be approximated as
\be{dudv}
 u'(v)= \alpha \left(\frac{u}{v} - 2\right)
,~~~\alpha =-\frac{w_1}{2w_2}. 
\ee
Solving the equation gives 
\be{origin}
u = 
\left\{
\begin{array}{cc}
\frac{2\alpha}{\alpha-1}v+ q \left(\frac{v}{v_b}\right)^{\alpha}~~~~~(\alpha\neq 1),
\\
\\
-2v\log |v|+q'~ v~~(\alpha=1),
\end{array}
\right.
\ee
where $q$ and $q'$ are integration constants.
Because we are interested only  in $w_1w_2<0$ case,
the constant $\alpha$ takes positive value 
($\alpha>0$)\footnote{It is to be noted that  when $\alpha  < 0$, 
Eq.~\eqref{origin} implies that 
the solution curve never passes the point $\mathcal{O}$ unless $q = 0$. 
Therefore, only linear behavior exists at $\mathcal{O}$. }.
The solution curve of a regular wormhole
should not end in the region $u\to -\infty$ or $v \to \infty$. 
Thus, both ends of a solution curve should be located at $\mathcal{O}$. 
In this sense, the point \oo plays the role of an asymptotic infinity.

In Eq.~\eqref{origin}, when $\alpha>1$,
the linear behavior dominates since $v^{\alpha} \ll 1$.
When $0< \alpha < 1$, the power law behavior (the $v^\alpha$ term) dominates. 
When $\alpha = 1$,  the solution takes $u \sim -2v \log |v| $ behavior.

The radius can be expressed in terms of $v$ as
$$
r = r_0 \left(\frac{v}{v_b}\right)^{-\alpha} ,
$$
where $r_0$ is a constant.
The mass function at $r$ becomes 
\be{m:O}
m(r) = \frac{u r}{2} =
	\frac{r_0}{2} \left[q-\frac{2\alpha}{1-\alpha}  \Big(\frac{r}{r_0}\Big)^{(\alpha-1)/\alpha}  \right] .
\ee
When $\alpha \geq 1$ ($0<w_2 \leq -w_1/2$), 
the mass function diverges and
the geometry may not  be flat asymptotically.
This mass function approaches a finite value $q r_0/2 $ asymptotically
only when $0< \alpha < 1$ ($-2w_2< w_1< 0$ or $0< w_1 < -2w_2$).

To see the $g_{tt}$ part of the metric around $\mathcal{O}$,
we need to consider the next order form of $r(v)$.
It can be calculated by integrating Eq.~\eqref{dr:u}
incorporating the relation Eq.~\eqref{origin} with $0<\alpha<1$:
\bea
r&=&r_0\left(\frac{v}{v_b}\right)^{-\alpha}
\exp{\left[\frac{(1+w_1)\alpha q}{2w_1} \left(\frac{v}{v_b}\right)^{\alpha}\right]},\nn\\
\left[\frac{v(r)}{v_b}\right]^\alpha&=&\frac{r_0}{r}-\frac{(1+w_1)\alpha q}{2w_1}\left(\frac{r_0}{r}\right)^2+\frac32\left(\frac{(1+w_1)\alpha q}{2w_1} \right)^2\left(\frac{r_0}{r}\right)^3+\cdots.
\eea
 Then, the $g_{tt}$ part of the metric around $\mathcal{O}$ becomes
\bea
f(r) &=& f_0 \left(\frac{r}{r_0}\right)^{\frac{4w_2}{1+w_1}} \left(\frac{v}{v_b}\right)^{-\frac{2w_1}{1+w_1}} \nn\\
&=& f_0\left(1-\frac{2m_\infty}{r}\right)+\mathcal{O}\left(\frac{1}{r^2}\right),
\eea
where we use Eq.~\eqref{f:r1} with the relation $v=4\pi r^2\rho$ 
and $m_\infty=r_0 q$ denotes the asymptotic value of mass in Eq.~\eqref{m:O}.
This guarantees asymptotic flatness for $0<\alpha<1$ case.

The density becomes
$$
\rho = \frac{v}{4\pi r^2} = \frac{1}{4\pi  r_0^2} 
		\left(\frac{r_0}{r}\right)^{2-2w_2/w_1} .
$$

\subsubsection{Behavior  around the wormhole throat  \PT}
	\label{sec:332}
Let us investigate  the behavior of a solution curve $C$ around the throat \PT$(1,v_b)$.
By using the trial function $1-u = \kappa |v-v_b|^\beta$ around $u\sim 1$, 
we find that Eq.~\eqref{de2} allows a quadratic and a linear behaviors for $C$,
\begin{eqnarray} \label{hor}
\mbox{i) The symmetric solution } &&~\beta =1,   \qquad u = 1 - s^{-1} ( v- v_b),  \label{hor2}\\
\mbox{ii) The asymmetric solution}&& ~\beta = 2,   \qquad  u =1- \kappa  (v-v_b)^2, 
\end{eqnarray}
where $\kappa$ denotes an arbitrary real number and $s$ is given in Eq.~\eqref{slope}.
Here `asymmetric'/`symmetric' implies that
the geometric structure is asymmetric/symmetric 
with respect to the inversion relative to the throat, respectively.

For the two cases, the radius takes the form,  
\be{r:P2_i}
r \approx b \left[1+ \frac{w_1}{1+w_1}(1-u) \right].
\ee
Therefore, $r$ takes minimum value at $v= v_b$ when $\kappa > 0$
for the case ii). 
As will be shown in the numerical plot, 
the symmetric solution appears as a large $\kappa$ limit of the asymmetric solution. 
The example of the exact symmetric solution
will be given in Eq.~\eqref{linear sol} when $1+w_1+2w_2 =0$ i.e. $s= -v_b$.

It appears that $u,v$ and $du/dv$ are continuous at \PT~even for solution curves 
with different $\kappa$.
Therefore one should check whether
solutions with different $\kappa$ can be attached smoothly
to form a proper wormhole solution or not.
To answer this question,
let us calculate the extrinsic curvatures.
As in Ref.~\cite{Israel:1966rt}
we divide a wormhole spacetime into upper/lower parts
with respect to the throat
and denote them $\Sigma_\pm$, respectively.
Then in our coordinates,
the extrinsic curvatures are 
\bea
K^\pm_{ab}
=\frac12\frac{\partial g^\pm_{ab}}{\partial r},~~~~a=t,\theta,\phi.
\eea
Since $K_{\theta\theta}$ and $K_{\phi\phi}$ have the same values on both sides, 
we consider $K_{tt}$ only here.
From the metric Eq.~\eqref{metric} and 
the form of $g_{tt}$ in Eq.~\eqref{f:r1},
we have
\bea
K_{tt}=-\frac12\frac{d f}{dr}=\left(
-\frac{2w_2}{1+w_1}\frac{1}{r}+\frac{w_1}{1+w_1}\frac{1}{v}\frac{d v}{dr}\right)f(r).
\eea
By using Eq.~\eqref{dr:u} and the above relations \eqref{hor},\eqref{hor2}, 
$K_{tt}$ behaves as
\bea
K_{tt}=\left\{
\begin{array}{cc}
\frac{1+w_1+2w_2}{1+w_1}\frac{f(b)}{b},
~~~~~~~\beta=1,
\vspace{.2cm}
\\
-\frac{w_1}{\kappa(v-v_b)}
\frac{f(b)}{b}, ~~~~~~~~\beta=2.
\end{array}
\right.
\eea
Thus for $\beta=1$ case,
two solutions can be attached smoothly.
However, for $\beta=2$ case,
the boundaries require a `surface layer'
if $K^+_{ab}\neq K^-_{ab}$.
In other words,
the upper/lower parts 
cannot be adjoined for different $\kappa$'s.

\subsubsection{Behavior around the repelling point \PB } \label{sec:PB}
The repelling point \PB~is where B2 meets R2.
In Ref.~\cite{Sorkin:1981wd},
this crossing point was shown to act as an `attractor' 
in the dynamics of self-gravitating ball of radiation
that a solution curve approaches \PB~in a spiral form.
On the other hand, in wormhole spacetimes,
solution curves rather avoid \PB~
than take the spiral behavior.
In this sense, the point \PB~acts as a `repeller'.

Let us approximate the differential equations~\eqref{de2} and~\eqref{dr:u} around \PB. 
After setting $u = u_B + x$ and $v= v_B + y$, they can be written as, 
to the linear order in $x$ and $y$,
\be{dxy/xi}
\frac{d}{d \log r}  \left( \begin{array}{c}
    x \\ 
    y \\ 
  \end{array} \right)=   \left(\begin{array}{cc}
    -1 & 2 \\ 
  (1-\gamma)s &~\gamma-1 \\ 
  \end{array}  \right)\left( \begin{array}{c}
    x \\ 
    y \\ 
  \end{array} \right) , \qquad  
 \gamma \equiv 1- \frac{2w_2}{1+w_1} = \frac32 +w_1 s > 0. 
\ee
Defining new variables
$
X_\pm =y -\frac{\gamma \pm \sqrt{\gamma^2 + 8s (1-\gamma) }}{2s(1-\gamma)}x ,
$
Eq.~\eqref{dxy/xi} is reduced to a diagonal form,
\be{X:asym}
\frac{dX_\pm}{d\log r} = \epsilon_\pm X_\pm; 
\qquad 
\epsilon_\pm 
	= \frac{-2+\gamma \mp \sqrt{\gamma^2 + 8s(1-\gamma) }}{2} .
\ee
Note that the term in the square-root is 
\bea
D \equiv \gamma^2 +8s(1-\gamma) 
=w_1(w_1-8)(s-s_+)(s-s_-),
\eea
where
\bea
s_\pm =\frac{4-3w_1\pm4\sqrt{1+3w_1}}{2w_1(w_1-8)}.
\eea

If $D<0$, the eigenvalues $\epsilon_\pm$ have imaginary parts
and the solution curve $C$ has a spiral behaviors asymptotically 
as depicted in Ref.~\cite{Sorkin:1981wd}.
In our case, this never happens.
The proof is the following:
\begin{itemize}
\item $w_1<-1$\\
Interpreting $D$ as a quadratic function of real number $s$, 
one can find that $D=0$ does not have a real valued solution.  
Hence $D$ is always positive because $w_1(w_1-8)>0$.
Therefore, when $w_1<-1$, one never gets a `spiral' solution.
\item  $w_1>0$. In this case,  $s_\pm$ are real numbers.
  \begin{enumerate}
   \item $0<w_1<8,~(s_+<s_-)$: 
In order for $D> 0$,  $s$ should satisfy $s_+< s < s_-$. 
As seen in Fig.~\ref{fig:cases2}, the point \PB~is located in a physical region only for the type V, 
which restricts $v_b \leq s\leq 0$. 
At the present case, $s_+< v_b $ and $s_-> 0$. Therefore, $D$ is positive definite. 

\item $w_1 = 8$: In this case, $D=0$ and the eigenvalues are real-valued with $\epsilon_+= \epsilon_-$.
 \item $ w_1>8,~(s_-<s_+)$:  In order for $D>0$,  $ s< s_-$ or $s> s_+$. 
Since physical region of $s$ is $(v_b,0)$, in order to avoid the `spiral' curve, 
$(s_-,s_+)\cap (v_b,0)$ needs to be an empty set.
One can show that $s_+<v_b$ for $w_1>8$.
Consequently, the `spiral' curve never happens.
\end{enumerate}
\end{itemize} 
Since we always have $D>0$, the curve $C$ in the $(X_+,X_-)$ coordinates satisfies
\bea
X_\pm=X_{\pm,0}\left(\frac{r}{r_R}\right)^{\epsilon_\pm}\rightarrow \frac{X_-}{X_+}
=\frac{X_{-,0}}{X_{+,0}}~r^{-\sqrt D},
\eea
where $X_{\pm,0}$ are integration constants.
In other words, as $r$ increases towards $r_R$, 
the curve changes gradually (not oscillating) to the limit $(X_{-,0}/X_{+,0})~r_R^{-\sqrt D}$.

This gives $X_+/ X_{+,0}= (X_-/X_{-,0})^{\epsilon_+/\epsilon_-} $. 
For types I, II and V, where $\epsilon_\pm$ are real, 
$$
\epsilon_+ \epsilon_-
=\frac{2w_2}{w_1} \left(1+\frac{4w_2}{(1+w_1)^2}\right) < 0.
$$
This implies that every solution curve approaches along one of the $X_\pm$ axis
and departs along the other axis without touching the point \PB. 
Note that for the type IV, 
the repelling point is located outside of 
the region we are interested in.
%

\section{Numerical solutions}\label{sec:numeric}
In this section, 
we present and display  numerical solutions of the Einstein's equation~\eqref{de2} 
on the $(u,v)$ plane to show behaviors of solutions.

\subsection{$w_1< -1$ case} 
A regular wormhole  made of an anisotropic matter 
with non-negative energy density exists only for the Type I, where $w_1< -1$ and $w_2 > -(1+w_1)/4$.
A well-localized solution having finite total mass exists when $w_2> -w_1/2$. 
Therefore, we restrict our interests only to this situation. 

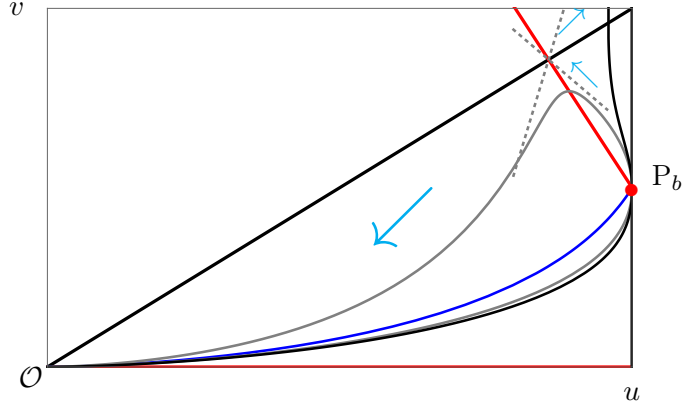
\begin{figure}[htb]
\begin{center}
\begin{tabular}{c}
\includegraphics[width=.5\linewidth,origin=tl]{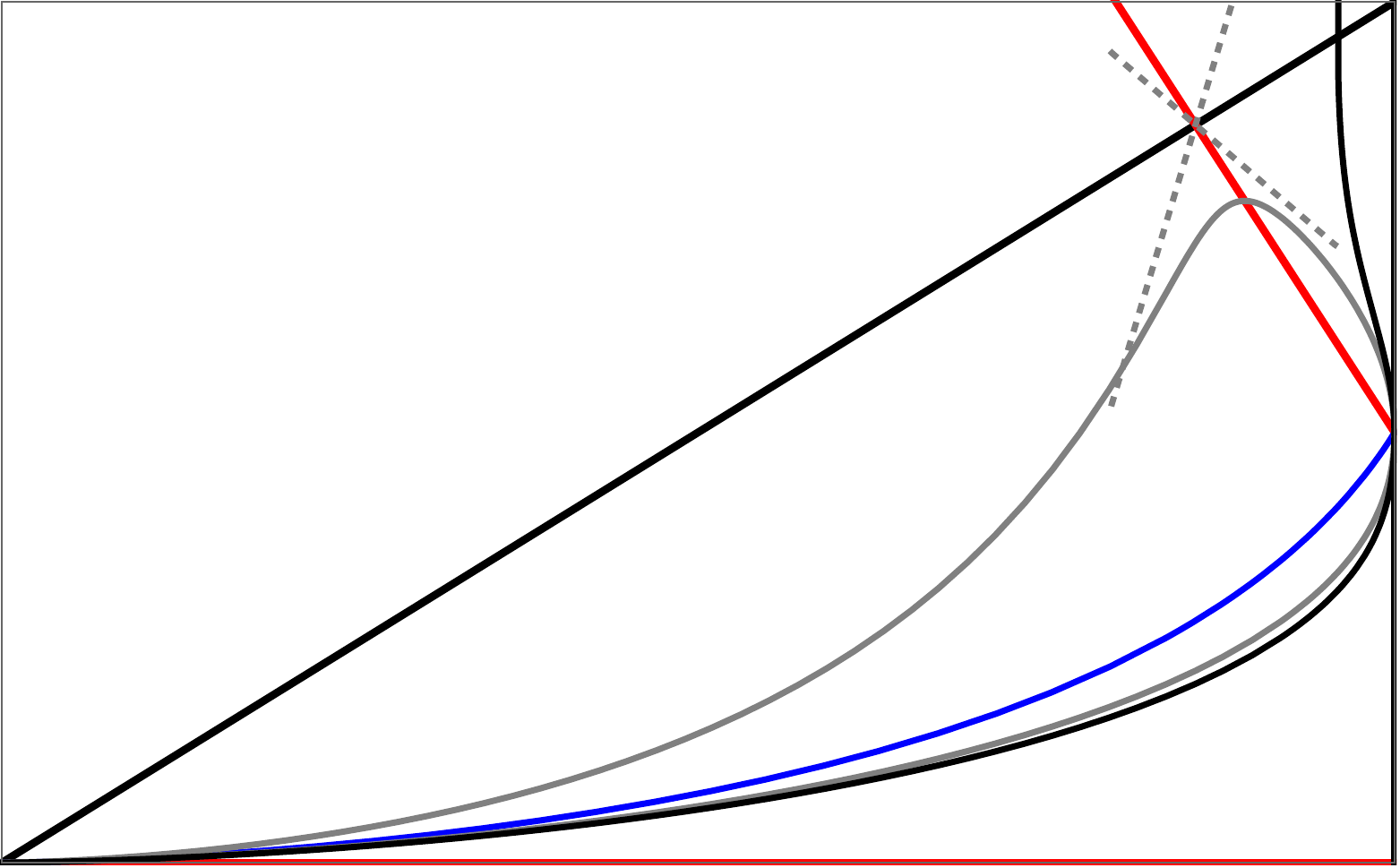}
\end{tabular}
\put (-10,-75) {$u$  }
\put (-240,70) {$v$  }
\put (-240,-70) { $\mathcal{O}$  }
\put (-3,6) { \PT} \put (-10,1) {\tcr{\large $\bullet$}}
\put (-35,65) {\textcolor{cyan}{  $\nearrow $  }}
\put (-30, 45) {\textcolor{cyan}{  $\nwarrow $  }}
\put (-105, -12) {\textcolor{cyan}{ \Huge $\swarrow $  }}
\end{center}
\caption{
Numerical solutions for the Type I with $w_1=-2$ and $w_2=1.5$.
}
\label{fig:asy}
\end{figure}
\begin{enumerate}
\item {\it Symmetric Solution:}
The blue curve in Fig.~\ref{fig:asy} corresponds 
to the linear behavior ii) at \PT~in Eq.~\eqref{hor2}.
The curve represents half of the wormhole solution from the throat at \PT\, 
to an asymptotic region at $\mathcal{O}$.
The other half of the wormhole can be added by duplicating the curve. 

\item {\it Asymmetric Solution:}
Given a positive $\kappa$ in Eq.~\eqref{hor}, 
an asymmetric solution curve is specified.
The gray curve represents a solution corresponding to a regular wormhole. 
In this case, the geometry of the wormhole is not symmetric with respect to the throat. 
Any curve which begins at \PT~and ends at \oo represents half of the spacetime 
including a throat and an asymptotic region. 
In the lower part of the gray curve,
 the value of $u$ monotonically decreases and forms an asymptotically flat region at $\mathcal{O}$.
 While, for the case of the solution curve corresponding to the other half, 
 $v$ bounces back around \PB~and starts to decrease. 
Then, it also forms the other asymptotically flat region at $\mathcal{O}$.

\item {\it Solution with one asymptotic region:}
Two ends of the black curve go into the region \oo and $v\to \infty$, respectively. 
The wormhole throat is located at \PT. 
Following the upper part of the black curve, the radius increases to a finite value as $v\to \infty$.
Therefore, a singularity of density exists.  

\end{enumerate}

\subsection{$w_1> 0$ case} 
In this case, the energy density is negative definite.
A regular wormhole solution exists only for the Type IV, where $w_2 < -(1+w_1)/4$.
A well localized solution having finite total mass exists when $0<w_1< -2w_2$. 
Therefore, we restrict our interests only to this situation. 

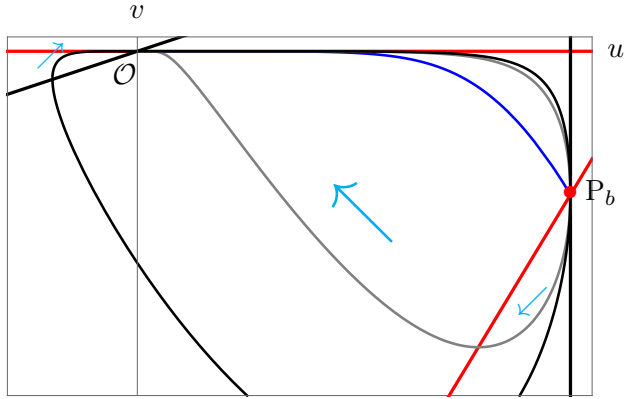
\begin{figure}[htb]
\begin{center}
\begin{tabular}{c}
\includegraphics[width=.5\linewidth,origin=tl]{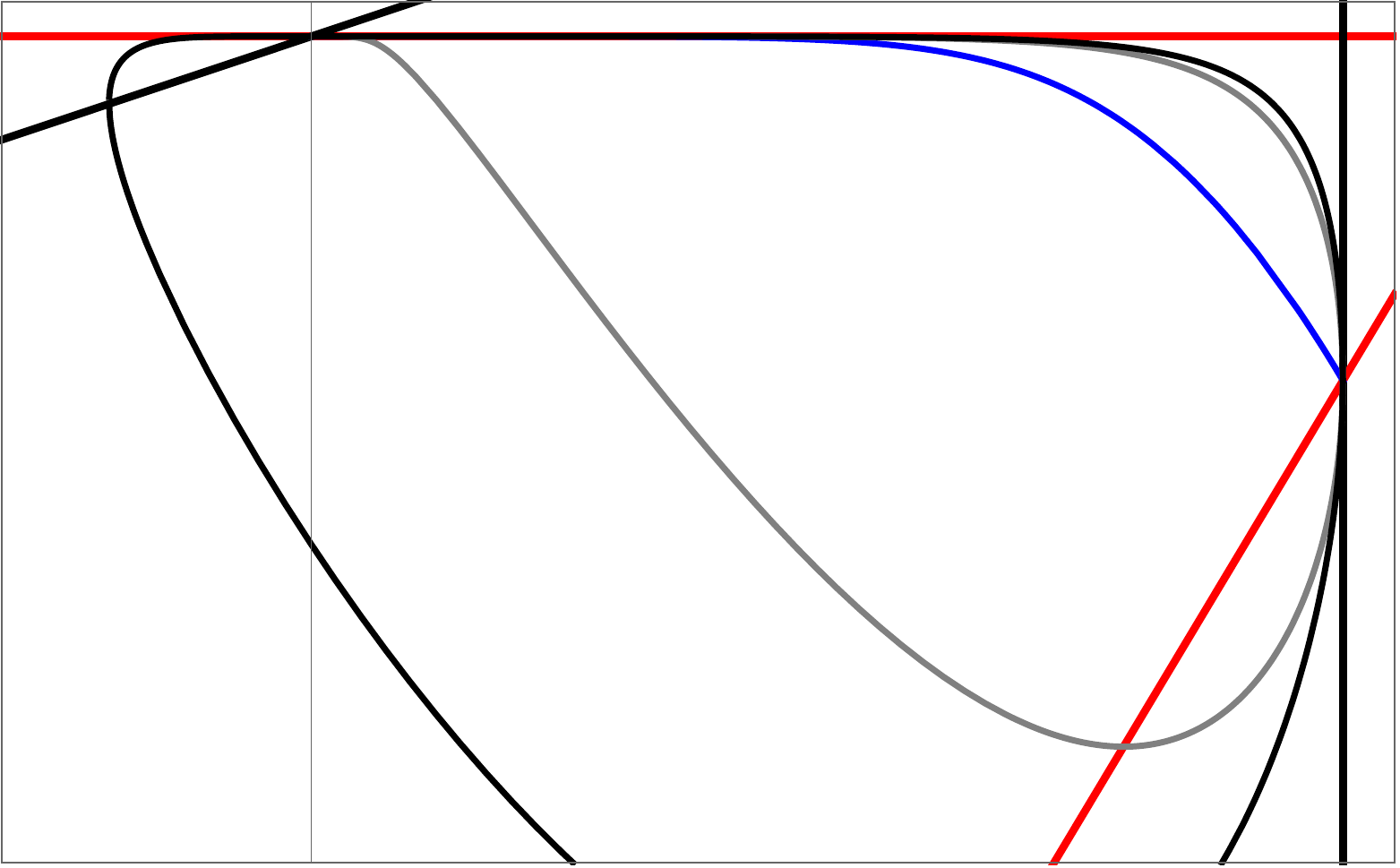}
\end{tabular}
\put (-1,65) {$u$  }
\put (-180,80) {$v$  }
\put (-190,55) { $\mathcal{O}$  }
\put (-13,11) { \PT} \put (-18,11) {\tcr{\large $\bullet$}}
\put (-215, 63) {\textcolor{cyan}{  $\nearrow $  }}
\put (-35,-30) {\textcolor{cyan}{\large  $\swarrow $  }}
\put (-105, 0) {\textcolor{cyan}{ \Huge $\nwarrow $  }}
\end{center}
\caption{
Numerical solutions for the type IV with $w_1=2$ and $w_2=-3$.
}
\label{fig:sol4}
\end{figure}

\begin{enumerate}
\item {\it Symmetric Solution:}
The blue curve in Fig.~\ref{fig:sol4} corresponds to the linear behavior ii) at \PT~in Eq.~\eqref{hor2}.
The curve represents half of the wormhole solution from the throat at \PT\, 
to an asymptotic region at $\mathcal{O}$.
The other half of the wormhole can be added by duplicating the curve. 

\item {\it Asymmetric Solution I:}
The gray curve presents a solution curve corresponding a regular wormhole in which
the mass function is positive definite in all the spacetime. 
In this case, the geometry is not symmetrical around the throat.

In the upper part of the gray curve,
 the value of $u$ monotonically increases and forms an asymptotically flat region at $\mathcal{O}$.
 While, for the case of the solution curve corresponding to the other half, 
 $u$ decreases at the beginning but bounces back and increases. 
Then, it also forms the other asymptotically flat region at $\mathcal{O}$.

\item {\it Asymmetric solution II:}
Two ends of the black curve go into the point $\mathcal{O}$.
 In this case, one end of $C$ approaches \oo starting from the negative value of $u$. 
 Therefore, the mass function becomes negative in this region.
Only the asymptotic behaviors around the point \oo are different from the first case. 
\end{enumerate}

\section{Analytic solutions}\label{sec:exact}

Even though one cannot find a general solution of Eq.~\eqref{TOV2}
it is still possible to find a few new analytic solutions.
In this section, we present two exact solutions for the cases $w_2 = -(1+w_1)/2$ and $w_2 = -(1+w_1)/4$.

\subsection{Various solutions for the case $1+ w_1+2w_2=0$ } \label{sec:IIIA}

First, we present an exact solution for the matter 
which marginally satisfies the strong energy condition $\rho + \sum_i p_i\geq 0$.
That is, when $\rho+\sum_i p_i=(1+w_1+2w_2)\rho=0$,
 Eq.~\eqref{TOV2} permits a solution.
 In this case, the slope $s = 1/2w_1 = -v_b$.
 
\begin{figure}[htb]
\begin{center}
\begin{tabular}{cc}
\includegraphics[width=.35\linewidth,origin=tl]{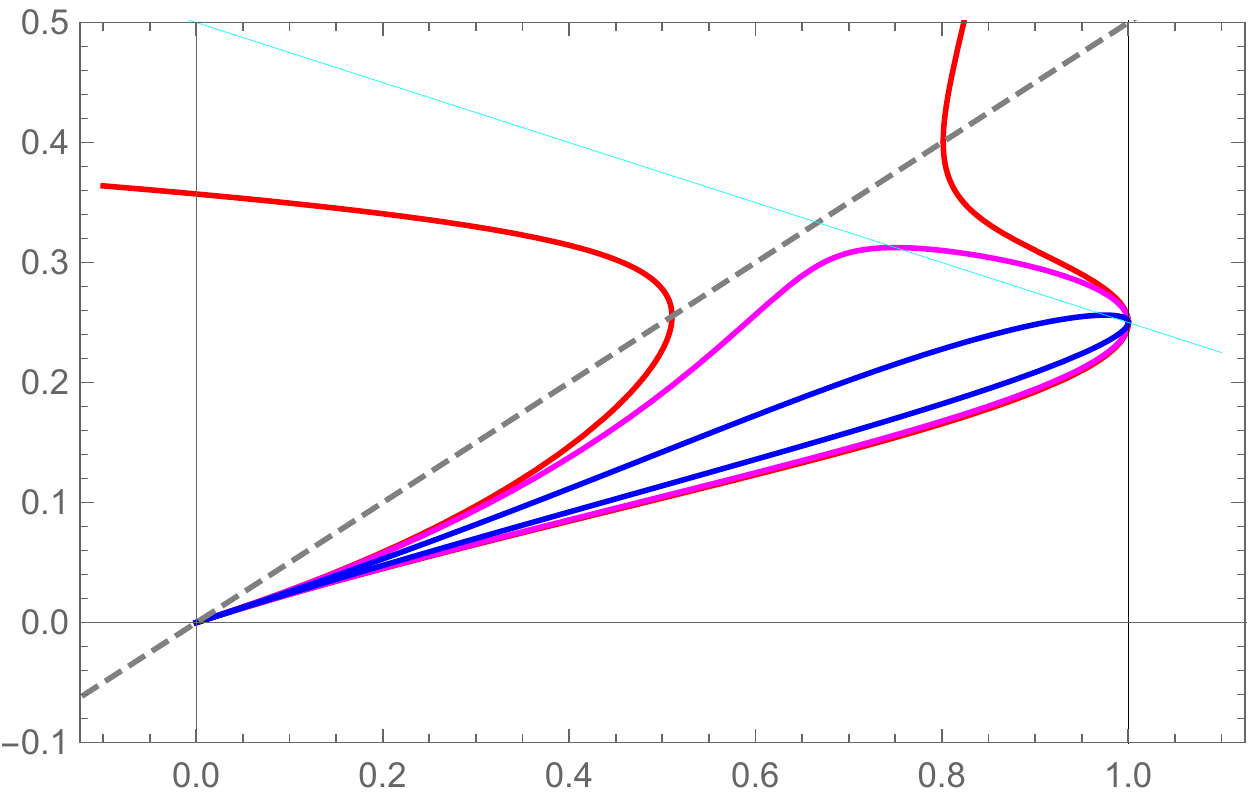} &~~
\includegraphics[width=.35\linewidth,origin=tl]{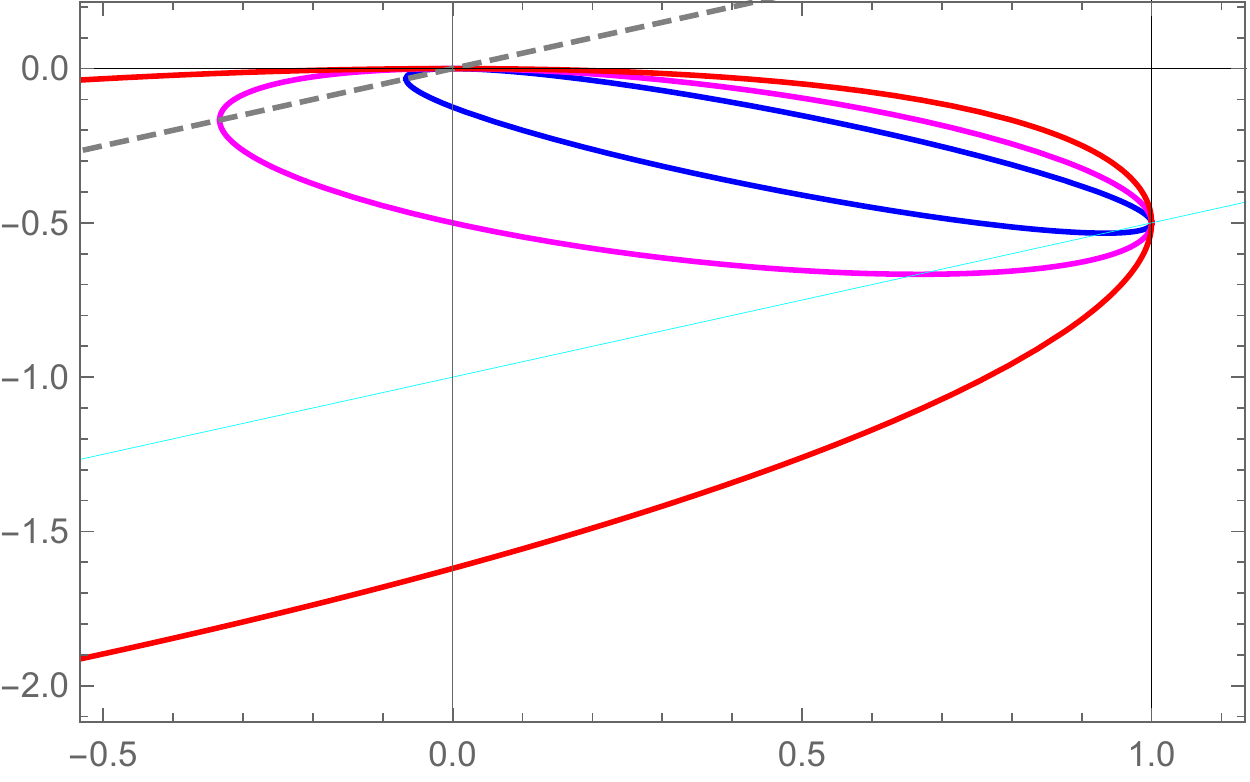} 
\end{tabular}
\put (-190,-32) {$u$  }
\put (-320,42) {$v$  }
\put (-325,-32) { $\mathcal{O}$  }
\put (-13,38) {$u$  }
\put (-108,53) {$v$  }
\put (-120,45) { $\mathcal{O}$  }
\end{center}
\caption{
Various solution curves  for $1+w_1+2w_2=0$.
(Left) The curves for $(w_1,w_2) =(-2,1/2)$
for various $\lambda=0.3,0.64,$ and 0.7 respectively for the blue, magenta and red curves;
(Right) The curves for $(w_1,w_2) =(1,-1)$.
Here $\lambda = 0.5,1$  and 1.8, respectively for the blue, magenta and red curves.
Every solution curves pass the point $(u,v) = (1, v_b)$ where a wormhole throat forms.
}
\label{fig:exactt0}
\end{figure}
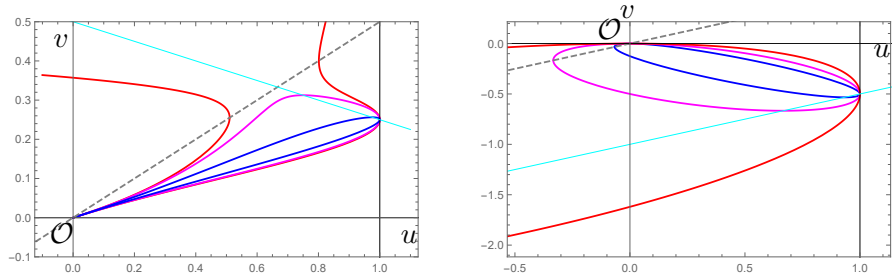

One can solve the TOV equation~\eqref{TOV2} in an explicit form:
\bea
\label{t0sol}
\left|\frac{v}{v_b}- u\right|\left(\frac{v}{v_b}\right)^{-\frac{w_1}{1+w_1}}
=\lambda \left(1-u\right)^{1/2},
\eea
where $\lambda$ is a positive constant.
One can confirm that this solution satisfies the autonomous equation~\eqref{de2}
for $1+w_1+2w_2=0$.
The solutions $v(u)$ are characterized by the constant $\lambda$.
We present various solutions $v(u)$ in Fig.~\ref{fig:exactt0}.
\\
Series expanding Eq.~\eqref{t0sol} around the throat,
gives $\kappa$:
\bea \kappa=(v_b \lambda)^{-2}.
\eea
Given a solution $v(u)$, one can obtain $u(r),v(r)(\equiv v(u(r))$ as functions of $r$
by using the Eq.~\eqref{dr:u}:
\bea
r=\exp\left(\int\frac{du}{2v(u)-u}\right).
\eea

One can readily check that,
when $C=0$, Eq.~\eqref{t0sol} becomes a linear solution curve
\be{linear sol}
v = v_b u .
\ee
In fact, one of the solution found in Ref.~\cite{Morris:1988cz} 
is a special case of the linear solution when $w_1=-2$.
This linear solution plays an important role in analyzing the solution space of Eq.~\eqref{de2}.

The radius and the density behaves as
\be{r,rho:v}
r = b \left(\frac{v}{v_b}\right)^{-\frac{w_1}{1+w_1}}, \qquad \rho
 = \frac{v}{4\pi r^2} = \frac{v_b}{4\pi b^2} \left(\frac{b}{r}\right)^{3+\frac{1}{w_1}} .
\ee
At the point $\mathcal{O}(0,0)$, the radius diverges to form a regular asymptotic region. 
At the point $r= b$, it forms a wormhole throat at $v= v_b$.
The mass function becomes
$$
m(r) = \frac{ru}{2} = \frac{b}{2} \left(\frac{r}{b}\right)^{-\frac{1}{w_1}} .
$$
For $w_1> 0$, the mass monotonically decreases from $b/2$ to zero as $r\to \infty$. 
Therefore, as $r\to\infty$,
 the total mass becomes divergent/zero 
when $w_1\lessgtr 0$.

The metric can be reconstructed to be
\bea
g_{tt}(r)&=&-f_0,
\nn\\
g_{rr}(r)&=&\frac{1}{1-\left(\frac{b}{r}\right)^{(1+w_1)/w_1}}.
\eea
In other words the temporal component of the metric is independent of $r$.
The radial component for the metric linearly diverges as $r \to b$, 
which is a signature of the wormhole throat. 

\subsection{Various solutions for the case $1+w_1 + 4w_2=0$} \label{sec:IIIB}

A new analytic wormhole solution exists in this case. 
The strong energy condition will be violated for $w_1<-1 $ because $1+w_1+2w_2 = -2w_2 < 0$. 
Since the slope $s=0$,
the denominator of Eq.~\eqref{de2} has no $u$ dependence
and the equation is integrable exactly to give
\be{sol1}
u = 1- \frac{\left(1-\frac{v}{v_b}\right)^2 }{ 1+\frac{2w_1}{1-w_1}\left( \frac{v}{v_b}\right) 
- c_1\frac{1+w_1}{1-w_1} \left(\frac{v}{v_b}\right)^{\frac{2w_1}{1+w_1}} },
\ee
where $c_1$ is an integration constant.
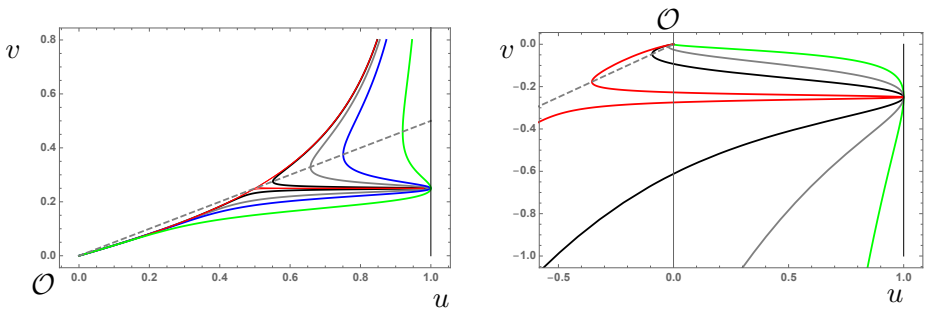
\begin{figure}[htb]
\begin{center}
\begin{tabular}{cc}
\includegraphics[width=.35\linewidth,origin=tl]{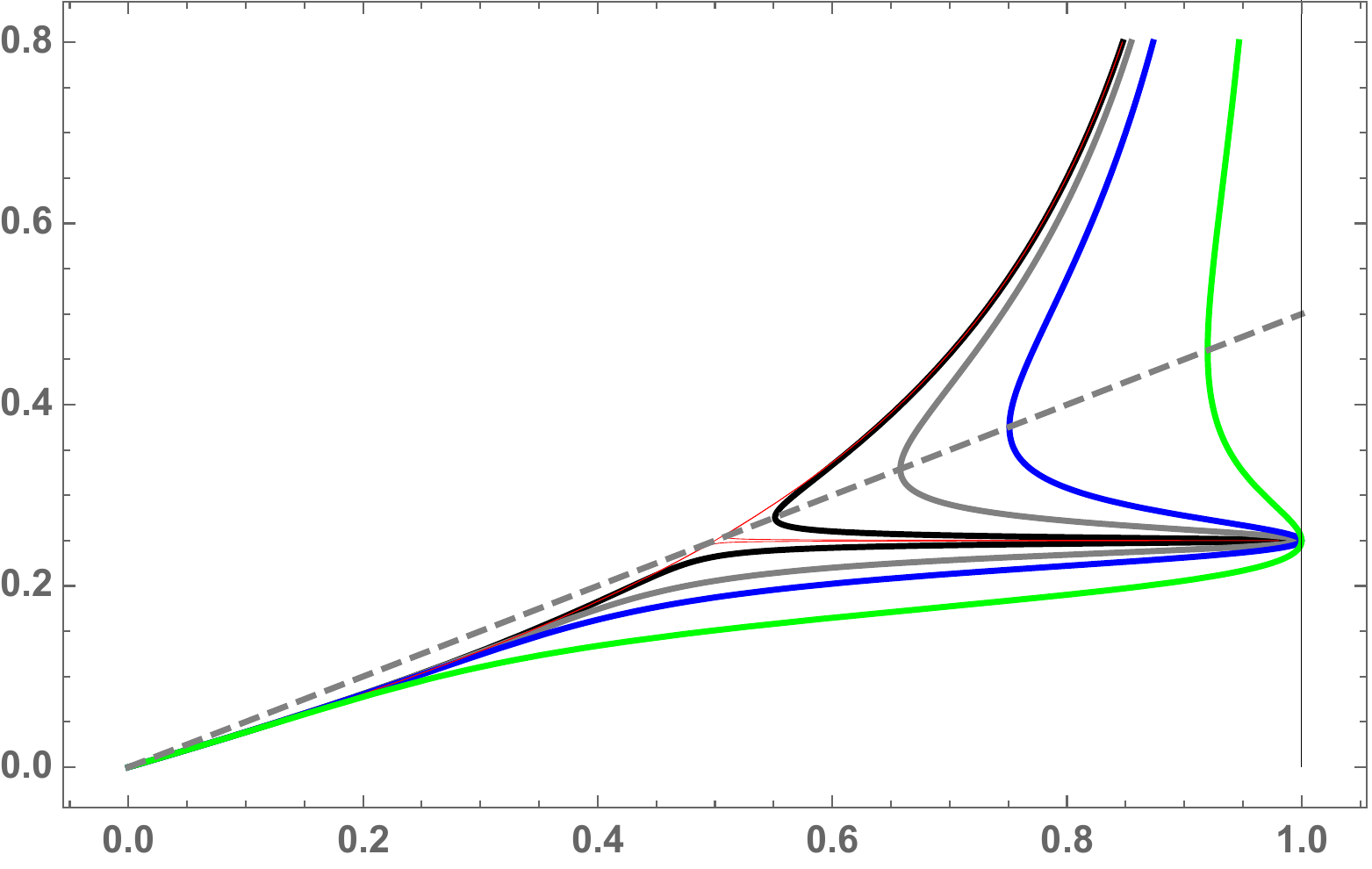} &~~
\includegraphics[width=.35\linewidth,origin=tl]{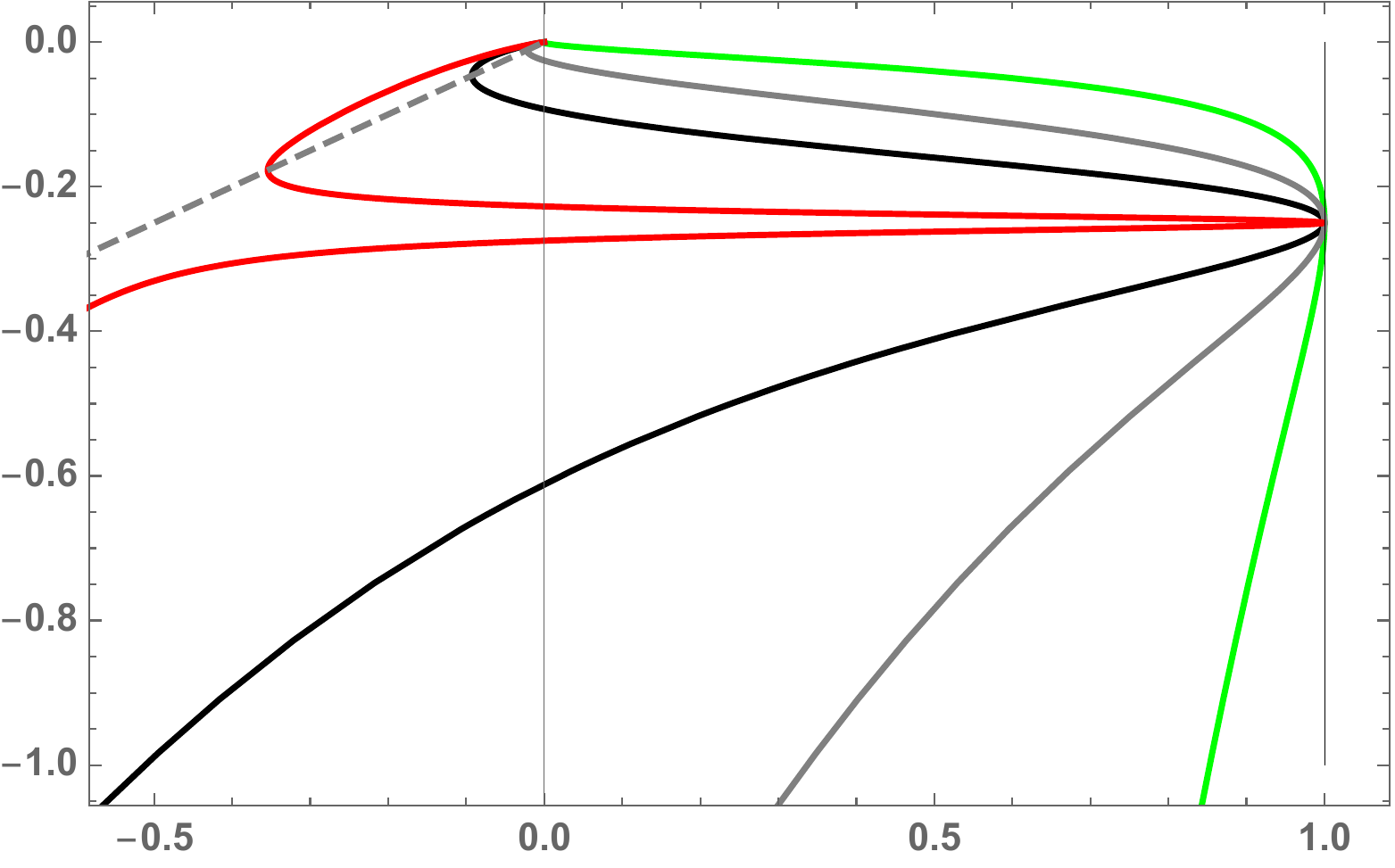} 
\end{tabular}
\put (-190,-52) {$u$  }
\put (-350,42) {$v$  }
\put (-344,-47) { $\mathcal{O}$  }
\put (-20,-50) {$u$  }
\put (-165,42) {$v$  }
\put (-110,53) { $\mathcal{O}$  }
\end{center}
\caption{
Various solution curves with $c_1> 1$ for $(w_1,w_2) =(-2,1/4)$ (Left) and $(2,-3/4)$ (Right).
Here $c_1 = 1.001, 1.1, 1.5, 2,$ and $ 4$, respectively for red, black, gray, blue, and green curves.
Every solution curves with $c_1 > 1$ pass the point $(u,v) = (1, v_b)$ where a wormhole throat forms.
}
\label{fig:exact}
\end{figure}

The characteristic behaviors of the solution curves with $c_1>1$ are given in Fig.~\ref{fig:exact} 
in which $w_1 = 2$ and $-2$, respectively for the left and the right panels. 
We neglected the curves with $c_1 \leq 1$ because they do not contain a wormhole throat. 
For both the cases, one end of the solution curves finishes at $\mathcal{O}$, 
which corresponds to an asymptotic infinity at $r \to \infty$. 
The other end finishes at $v\to \infty$ or $(u,v) \to (-\infty,-\infty)$, 
respectively for $w_1< -1$ or $w_1> 0$, where a singularity forms.  

Let us observe the behaviors of the solutions in detail.
Around $v = v_b$, it becomes, for $c_1 \geq 1$, 
$$
u \approx 1+ \frac{4w_1^2}{(c_1-1)}\frac{1-w_1}{1+w_1}( v -v_b)^2 , \quad c_1 > 1;
\qquad u \approx - \frac1{w_1} -\frac{2 (v/v_b -1)}{3w_1}  , \quad c_1 =1.
$$ 
Therefore, Eq.~\eqref{sol1} presents a relevant wormhole throat 
when $c_1> 1$ because $u\leq 1$ in this case. 
On the other hand, when $c_1=1$, 
the curve does not pass the point $(u,v) = (1,v_b)$, the throat.
Incidentally, the value of 
$\kappa$ in Eq.~\eqref{hor} is related with $c_1$ by 
$$
\kappa = \frac{4w_1^2}{1-c_1}\frac{1-w_1}{1+w_1} .
$$ 

Around $\mathcal{O}$, the asymptotic region, $u$ behaves as
\be{u:O}
u \approx 
\left\{
\begin{array}{cc}
\frac{2}{1-w_1} \frac{v}{v_b},~~~~~~~~~~~~~ &~~~~~ |w_1| >1,  \\
-c_1\frac{1+w_1}{1-w_1}  \left(\frac{v}{v_b}\right)^{\frac{2w_1}{1+w_1}} , 
&~~~~~ 0< w_1<1, \\
\frac{v}{v_b} \log \frac{c_1v}{v_b},~~~~~~~~~~~ & ~~~~~w_1 = 1.
\end{array}
\right.
\ee
When $|w_1|> 1$, the linear behavior dominates.
On the other hand, when $0< w_1 < 1$, polynomial behavior dominates.

The radius can be obtained  by using Eqs.~\eqref{de2} and \eqref{dr:u}:
\be{r:v1}
r = b \left(\frac{v}{v_b}\right)^{-\frac{2w_1}{1+w_1}}
	\frac{1-w_1}{(1+w_1)(1-c_1)} \left( 1+ \frac{2w_1}{1-w_1} \frac{v}{v_b} 
	-  c_1\frac{1+w_1}{1-w_1}  \Big(\frac{v}{v_b}\Big)^{\frac{2w_1}{1+w_1}}\right) .
\ee
The radius takes its minimum value $b$ at the throat $v=v_b$. 
As $v$ decreases, the radius goes to infinity at $v=0$. 
On the other hand, as $v \to \infty$ 
the radius increases to a local maximum value $bc_1/(c_1-1)$ or 
diverges with the form $r\propto v^{(1-w_1)/(1+w_1)}$,
 respectively for $|w_1|> 1$ or $ 0 < w_1 < 1$.

The mass function behaves as
\be{m:v}
m = \frac{u r}2 = \frac{b}{(1-c_1)(1+w_1)}
	\left(\frac{v}{v_b}\right)^{\frac{1-w_1}{1+w_1}} \,
\left[ 1-\frac{1-w_1}2 \frac{v}{v_b}
	- c_1\frac{(1+w_1) }2 \Big(\frac{v}{v_b}\Big)^{
		-\frac{1-w_1}{1+w_1}} \right].
\ee
For $w_1< -1$, the mass function has a divergent value for large $r$ ($v\to 0$).
This is a signature of the instability of the wormhole throat. 
The mass decreases to $m(v_b) = b/2$ and then bounces back to increase
 to a finite value $m(v\to\infty) = bc_1/2(c_1-1) $. 
For $w_1> 1$, the mass diverges  both at $v\to 0$ and as  $v\to \infty$.
On the other hand, for $0< w_1<1$, it goes to a constant value $m(v=0) = b c_1/2(c_1-1)$
at $v=0$
 but diverges as $v\to \infty$.

The $g_{tt}(\equiv -f$) part  of the metric function from Eq.~\eqref{f:r1} becomes
\be{f:r}
f(r) = f_0 \left(\frac{r}{r_0}\right)^{\frac{4(w_2-w_1)}{1+w_1}}
	 \left(\frac{\rho}{\rho_0} \right)^{-\frac{2w_1}{1+w_1}} 
    =f_0 \left(\frac{r}{r_0}\right)^{\frac{4w_2}{1+w_1}}
	 \left(\frac{v(r)}{v_0} \right)^{-\frac{2w_1}{1+w_1}}.
\ee
One may also check the two formula for $f(r)$ in Eqs.~\eqref{fr:g} and \eqref{f:r1} present the same result. 
The $g_{rr}$ part of the metric function is
\bea
g_{rr}(r)= 
\frac{ 1+\frac{2w_1}{1-w_1}\left( \frac{v(r)}{v_b}\right) 
- c_1\frac{1+w_1}{1-w_1} \left(\frac{v(r)}{v_b}\right)^{\frac{2w_1}{1+w_1}} }
{\left(1-\frac{v(r)}{v_b}\right)^2 }.
\eea
\vspace{.3cm}
Although the two exact solutions discussed in this section have deficits of their own,
they provide good insights into the general solution space. 

\section{Limiting solution for $|s| \gg 1$}
\label{sec:slimit}

One can rewrite the TOV equation~\eqref{de2}  as
\bea
\frac{du}{dv} =\frac{-1}{1+w_1}\left(\frac{2v-u}{v}\right)\frac{1-u}{v-su-v_s},~~~~~~v_s\equiv v_b-s =\frac{2w_2}{w_1(1+w_1)}.
\eea
One can see that when the value of $s$ is large, the value of $v_s$ is also large.
$v_s$ can be rewritten as
\bea
v_s\equiv -a s,~~~a\equiv\frac{4w_2}{1+w_1+4w_2}.
\eea
Now, the TOV equation can be expanded as a series of $s^{-1}$
\bea
\frac{du}{dv} &=& \frac{-1}{1+w_1}\frac{(2v-u)(1-u)}{v[v-s(u-a)]}\nn\\
&=&-\alpha\left(\frac{2v-u}{v}\cdot\frac{1-u}{1-u/a}\right)\left[1+\frac{v}{a-u}s^{-1}+\cdots \right],
\eea
where $\alpha=1/(1+w_1)as=-w_1/2w_2$ was defined  in Eq.~\eqref{origin}.
The general behavior of the $0^{\rm th}$ order solution is the same
as that of the solution around \oo given in Eq.~\eqref{origin}.
\be{sol:Ls}
u = -\frac{2\alpha}{1-\alpha} v +q \left(\frac{v}{v_b}\right)^{\alpha}, 
\ee
where $q$ is an integration constant.
For a solution curve passing the point $\mathcal{O}$, 
we should take $\alpha >0$ because  $v^\alpha\to\infty$ when $\alpha<0$.
Note that when $\alpha$ is given, the trajectories are classified by $q$ only.

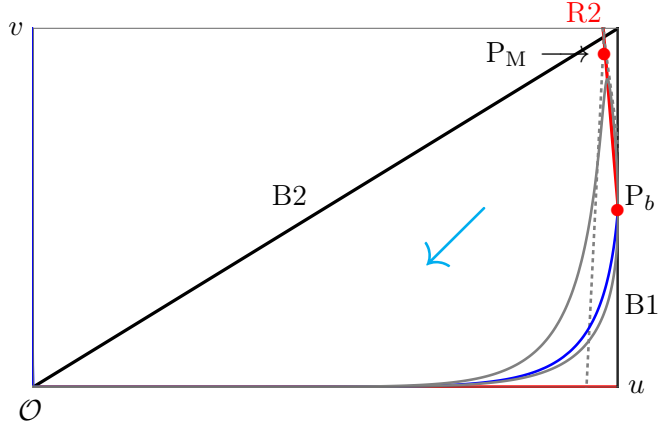
\begin{figure}[htb]
\begin{center}
\begin{tabular}{c}
\includegraphics[width=.5\linewidth,origin=tl]{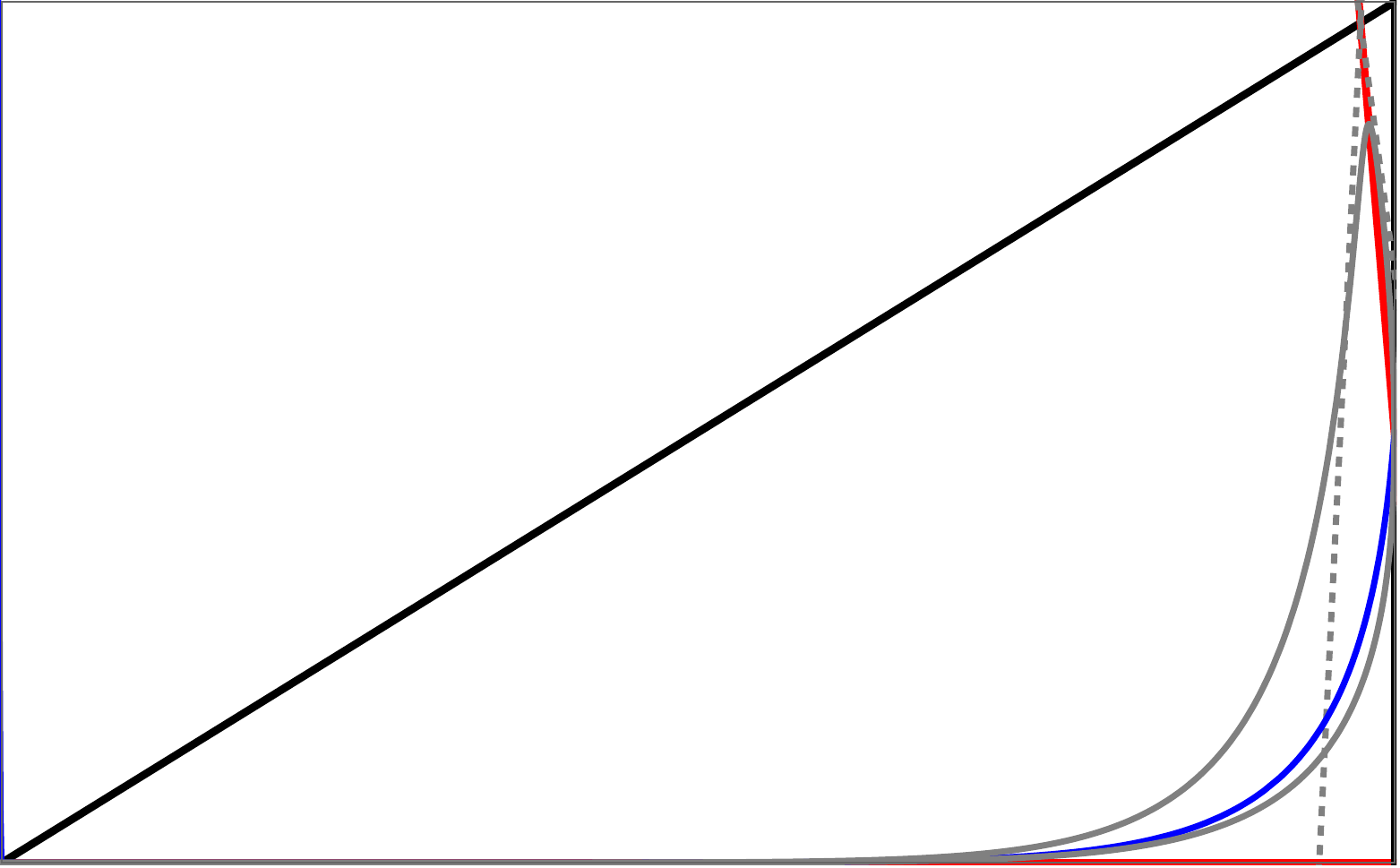}
\end{tabular}
\put (-3,-65) {$u$  }
\put (-235,70) {$v$  }
\put (-235,-75) { $\mathcal{O}$  }
\put (-30,75) { \tcr{R2}} \put (-15,60) {\tcr{\large $\bullet$}}
\put (-140,6) { B2} 
\put (-8,-35) { B1} 
\put (-60,60) { \PM}
\put (-8,6) { \PT} \put (-10,1) {\tcr{\large $\bullet$}}
\put (-80, -12) {\textcolor{cyan}{ \Huge $\swarrow $  }}
\put (-40,60) { \large$\longrightarrow$}
\end{center}
\caption{
Characteristic picture of limiting solution of Type I with $w_1< -1$.
}
\label{fig:limiting}
\end{figure}
In Fig.~\ref{fig:limiting}, characteristic picture of the limiting solution is given. 
The blue curve represents a regular wormhole solution which is symmetric with respect to the throat.
As seen in the figure, the curve gradually decreases to \oo with $r$.

 \vspace{1cm}
The gray curve in Fig.~\ref{fig:limiting} represents an asymmetric wormhole solution 
which is regular over the whole spacetime when $w_1<-1$.
Considering the upper half of the curve starting from the point \PT, 
the value of $v$ increases to 
a maximum value at \PM~and then decreases to zero 
as one approaches the asymptotic region. 
Let the coordinates of \PM~be $(u_M,v_M)$.
Then we have
\bea
\left.\frac{dv}{du}\right|_M=0, ~~~~v_M=s(u_M-a), ~~~a\equiv\frac{4w_2}{1+w_1+4w_2}.
\eea
Let us introduce a new coordinates $(x,y)$ around \PM~as 
$u\equiv u_M+x,v\equiv v_M+y$.
Then Eq.~\eqref{de2} can be approximated
for small $|x|,|y| \ll 1$ and $|y|\ll |sx|$ to be
\bea
\label{sol:max}
\frac{dy}{dx}=- c_M \left[x+\left(\frac{1}{1-u_M}+\frac{1}{2v_M-u_M}\right)x^2+\cdots \right] \quad \Rightarrow \quad
y=-\frac12 c_M x^2+\mathcal{O}(x^3),
\eea
where 
\bea
\label{cm}
c_M=\left|\frac{(1+w_1)s v_M}{(1-u_M)(2v_M-u_M)}\right|=\left|\frac{a}{\alpha}\frac{ v_M}{(1-u_M)(2v_M-u_M)}\right|.
\eea
The solution curve has approximately quadratic form around \PM, $y=-\frac12 c_M x^2$.
As one moves farther from the maximum point,
the curve takes higher order corrections.

When $|s|$ becomes large,
the line R2 approaches the line B1,
$a\to 1$ and $1-u_M\to 0$.
For the asymmetric solution,
the maximum point \PM ~goes
 near the repelling point \PB~($2v_M- u_M\to 0$).
That is, $c_M \to \infty$.
This means the solution curve becomes very steep
 and narrow as $|s|\to \infty$ 
as shown in Fig.~\ref{fig:limiting}.
Hence, in the limit,
the derivative of an asymmetric curve 
is discontinuous at \PM.
As $u_M\to 1$,
the curve mostly can be divided into
the left part and the right part of \PM.

Solution curves around \oo 
  are under the line B2 $(v<u/2)$.
Thus to meet \PM~or \PT,
derivative of the solution
takes only positive value:
\bea
\frac{dv}{du}=\frac{1}{\frac{2\alpha}{\alpha-1}+\alpha q\left(\frac{v}{v_b}\right)^{\alpha}}=\frac{v}{\alpha(u-2v)}>0,
\eea
for positive $v$.
Therefore the type of solution \eqref{sol:Ls} 
either can be extended to meet symmetric solutions (to \PT)
or to meet the left part of the asymmetric solutions (to \PM)
depending on the values of $\alpha$ and $q$.

\section{Summary and Discussions}\label{sec:summary}

We have classified static spherically symmetric wormholes
consisting of an anisotropic matter throughout the entire spacetime
and studied necessary conditions for the spacetime to be nonsingular.
In the process, we found a few exact solutions in special configurations.
We also presented wide variety of solutions 
by studying the properties they generally have.
The behavior of wormhole geometry and physical quantities around important points
such as the throat, the repelling point
and the asymptotes have been studied.
Numerical solutions have also been presented 
to give insights into the behaviors of general solutions.

We found that the throat geometry is insensitive to the angular pressure.
The derivative of the mass function depends solely on the radial pressure as in Eq.~\eqref{vT}.
The regular asymptotic geometry is determined by
the ratio of the radial to the angular pressures.

An anisotropic matter with the limit $w_1\equiv p_1/\rho\to -1$
resembles the stress-energy tensor of radial electric field
which may be used to produce a charged wormhole in Ref.~\cite{Kim:2001ri}.
This corresponds to the $|s|\to \infty$ limit 
which we analyzed in Sec.~\ref{sec:slimit}.
From the asymptotic form of the solution in Eq.~\eqref{sol:Ls},
the part of the solution near the maximum $v_M$
in Eq.~\eqref{sol:max}
and the numerical solution,
one can figure out
how the solution behaves.
However, we observed that 
there is no wormhole solution when $w_1=-1$.
One may wonder whether this is contradictory to the case that 
corresponds to a charged wormhole solution,
for example, that in Ref.~\cite{Kim:2001ri}.
There, 
the energy density and the pressure are given by
\bea
\label{chargedwh}
\rho &=&\rho^{(0)}+\rho^{(1)},\nn\\
p_1 &=&p_1^{(0)}+p_1^{(1)},
\eea
where $(\rho^{(0)}, p_1^{(0)})$ are 
those of non-charged wormhole solution
and $(\rho^{(1)}, p_1^{(1)})$ are the charged dressing, 
\bea
\rho^{(1)}=-p_1^{(1)}=\frac{Q^2}{8\pi r^4},
\eea
where $Q$ is the electric charge.
As one can see in the above equation~\eqref{chargedwh},
$p_1\neq -\rho$ or $w_1\neq -1$
unless $\rho^{(0)}=p_1^{(0)}=0$.
As the authors pointed out in the paper,
it corresponds to the $b(r)=0$ case in the reference
and the spacetime becomes the Reissner-Nordstr\"{o}m solution.

One may also use our solutions to construct a new solution which use exotic matter minimally. 
In the vicinity of the throat, 
the truncated solution of the present work
 can be matched with a Schwarzschild solution outside 
with appropriate junction conditions~\cite{Israel:1966rt}.
Stability of wormholes has also been studied in part
in the context of general relativity recently~\cite{Bronnikov:2013coa,Novikov:2012uj}.
Though the studies are not yet complete,
it was shown that the Ellis type wormhole can be stable.
As the stability of a black hole made physicist to consider black hole seriously,
we expect the stability of a wormhole does the role.

Most studies of wormhole concentrate on finding a solution of the gravitational field equations.
Recently the exact form of entropy function $S(r)$
for a self-gravitating anisotropic matter
was obtained~\cite{Kim:2019ygw}.
When a wormhole is made of an anisotropic matter
of radius $R$ and the throat radius is $B$,
the entropy is given by a sum of 
$S_1(R)-S_1(B)$ in one side
and $S_2(R)-S_2(B)$ in the other side of the wormhole.
Though there have been a few studies, the wormhole thermodynamics
should be addressed  to have more clear picture.

In this work, we cannot avoid 
the use of exotic matter because we are based on the general relativity.
The extension of the analysis to a modified theory of gravity is admirable.

\acknowledgments

This work was supported by the National Research Foundation of Korea grants funded by the Korea government NRF-2017R1A2B4008513.



\end{document}